\documentclass[journal=jctcce,manuscript=article]{achemso}
\setkeys{acs}{articletitle = true}
\usepackage{caption}
\usepackage{bpchem} % Chemical compounds
\usepackage{setspace}
\usepackage{fullpage}
\usepackage{graphicx}
\usepackage{epstopdf}
\usepackage{xspace}
\usepackage{booktabs}
\usepackage{pdflscape}
\usepackage{tablefootnote}
\usepackage[sort&compress,numbers,super]{natbib}
\usepackage{amsmath}
\usepackage{subcaption}
\usepackage{nameref}
\usepackage{xfrac}
\usepackage{multirow}
\usepackage[title]{appendix}
\usepackage[capitalize]{cleveref}
\usepackage[symbol,perpage]{footmisc}

\newcommand{\citeboth}[1]{\citeauthor{#1}\cite{#1}\xspace}
\newcommand*{\citen}{}% generate error, if `\citen` is already in use
\DeclareRobustCommand*{\citen}[1]{%
  \begingroup
    \romannumeral-`\x % remove space at the beginning of \setcitestyle
    \setcitestyle{numbers}%
    ref. \cite{#1}%
  \endgroup
}
\newcommand*{\citens}{}% generate error, if `\citen` is already in use
\DeclareRobustCommand*{\citens}[1]{%
  \begingroup
    \romannumeral-`\x % remove space at the beginning of \setcitestyle
    \setcitestyle{numbers}%
    refs. \cite{#1}%
  \endgroup
}

\usepackage{letltxmacro}
\LetLtxMacro{\originaleqref}{\eqref}
\renewcommand{\eqref}{eq.~\originaleqref}

\newcommand{\figref}[1]{Figure~\ref{#1}}
\newcommand{\tabref}[1]{Table~\ref{#1}}
\newcommand{\secref}[1]{Section~\ref{#1}}
\newcommand{\appendixref}[1]{Appendix~\ref{#1}}

\newcommand{\si}{Supporting Information\xspace}

\newcommand{\isoff}{Iso-Iso FF\xspace}
\newcommand{\isaff}{Aniso-Iso FF\xspace}
\newcommand{\isoffcc}{Iso-Iso-CC FF\xspace}
\newcommand{\isaffcc}{Aniso-Iso-CC FF\xspace}
\newcommand{\isaffold}{Slater-ISA FF\xspace}
\newcommand{\mastiff}{MASTIFF\xspace}
\newcommand{\anisoff}{Aniso-Aniso FF\xspace}

\newcommand{\sapt}{DFT-SAPT (PBE0/AC)\xspace}

\newcommand{\avtzm}{aug-cc-pVTZ+m\xspace}
\newcommand{\A}{\ensuremath{A_{ij}}\xspace}
\newcommand{\B}{\ensuremath{B_{ij}}\xspace}
\newcommand{\C}{\ensuremath{C_{ij,2n}}\xspace}
\newcommand{\R}{\ensuremath{r_{ij}}\xspace}

\newcommand{\fij}{\ensuremath{f(r_{ij})}\xspace}
\newcommand{\gij}{\ensuremath{g(\theta_i,\phi_i,\theta_j,\phi_j)}\xspace}
\newcommand{\sfunc}{\ensuremath{\bar{S}\text{-functions}}\xspace}

\newcommand{\dhf}{\ensuremath{\delta^{\text{HF}}}\xspace}

\newcommand{\Aex}[1]{\ensuremath{A^{\text{exch}}_{#1}}\xspace}
\newcommand{\Ael}[1]{\ensuremath{A^{\text{elst}}_{#1}}\xspace}

\newcommand{\Aind}[1]{\ensuremath{A^{\text{ind}}_{#1}}\xspace} % AJM Not ind,sr !!!
\newcommand{\Adhf}[1]{\ensuremath{A^{\dhf}_{#1}}\xspace} % AJM Not ind,sr !!!
\newcommand{\Adisp}[1]{\ensuremath{A^{\text{disp}}_{#1}}\xspace} % AJM Not ind,sr !!!

\newcommand{\aex}{\ensuremath{a_{i,lk}^{\text{exch}}}\xspace}

\newcommand{\erep}{\ensuremath{E^{\text{exch}}}\xspace}
\newcommand{\eelst}{\ensuremath{E^{\text{elst}}}\xspace}
\newcommand{\eind}{\ensuremath{E^{\text{ind}}}\xspace}
\newcommand{\edhf}{\ensuremath{E^{\dhf}}\xspace}
\newcommand{\edisp}{\ensuremath{E^{\text{disp}}}\xspace}

\newcommand{\vtot}{\ensuremath{V_{\text{FF}}}\xspace}
\newcommand{\vrep}{\ensuremath{V^{\text{exch}}}\xspace}

\newcommand{\velst}{\ensuremath{V^{\text{elst}}}\xspace}
\newcommand{\vind}{\ensuremath{V^{\text{ind}}}\xspace}
\newcommand{\vdhf}{\ensuremath{V^{\dhf}}\xspace}
\newcommand{\vdisp}{\ensuremath{V^{\text{disp}}}\xspace}

\newcommand{\vmultipole}{\ensuremath{\sum\limits_{tu}Q_t^iT^{ij}_{tu}Q_u^j}\xspace}
\newcommand{\vdrude}{\ensuremath{V_{\text{pol}}}\xspace}
\newcommand{\vdrudeind}{\ensuremath{V_{\text{pol}}^{(2)}}\xspace}
\newcommand{\vdrudescf}{\ensuremath{V_{\text{pol}}^{(3-\infty)}}\xspace}

\newcommand{\rmse}{RMSE\xspace}
\newcommand{\mse}{\ensuremath{\lVert\text{MSE}\rVert}\xspace}

\newcommand{\co}{\BPChem{CO\_2}\xspace}
\newcommand{\ho}{\BPChem{H\_2O}\xspace}
\newcommand{\nh}{\BPChem{NH\_3}\xspace}
\newcommand{\cl}{\BPChem{CH\_3Cl}\xspace}
\newcommand{\deltahsub}{\ensuremath{\Delta H_{\text{sub}}}\xspace}
\newcommand{\deltahvap}{\ensuremath{\Delta H_{\text{vap}}}\xspace}
\newcommand{\deltah}{\ensuremath{\Delta H}\xspace}
\newcommand{\kjmol}{\xspace\ensuremath{\text{kJ mol}^{-1}}\xspace}

\newcommand{\aniso}[1]{\ensuremath{\xi_i^{\rm{#1}}(\theta_i,\phi_i)}\xspace}

\author{Mary J. Van Vleet}
\affiliation[UW-Madison]
{Theoretical Chemistry Institute and Department of Chemistry, University of
Wisconsin-Madison, Madison, Wisconsin, 53706, United States}
\author{Alston J. Misquitta}
\affiliation[Queen Mary]
{Department of Physics and Astronomy, Queen Mary University of London, London E1 4NS, United Kingdom}
\author{J.R. Schmidt}
\email{schmidt@chem.wisc.edu}
\affiliation[UW-Madison]
{Theoretical Chemistry Institute and Department of Chemistry, University of
Wisconsin-Madison, Madison, Wisconsin, 53706, United States}

\title{New angles on standard force fields: towards a general approach for
treating atomic-level anisotropy}

\begin{document}
\maketitle
\onehalfspacing

%%%%%%%%%%%%%%%%%%%%%%%%%%%%%%%%% Abstract %%%%%%%%%%%%%%%%%%%%%%%%%%%%%%%%%%%%%%%%
\begin{abstract}

Nearly all standard force fields employ the `sum-of-spheres' approximation,
which models intermolecular interactions 
purely in terms of interatomic distances. Nonetheless,
atoms in molecules can have significantly non-spherical shapes,
leading to interatomic interaction energies with strong orientation
dependencies.
Neglecting this `atomic-level
anisotropy' can lead to significant errors in predicting interaction energies.
Herein we propose a simple, transferable, and computationally-efficient model
(\mastiff) whereby atomic-level orientation dependence can be incorporated
into ab initio intermolecular force fields. \mastiff includes anisotropic
exchange-repulsion, charge
penetration, and dispersion effects, in conjunction with a standard
treatment of anisotropic
long-range (multipolar) electrostatics.
To validate our approach, we benchmark \mastiff against various sum-of-spheres models
over a large library of intermolecular interactions between small organic
molecules. \mastiff
achieves quantitative accuracy with
respect to both high-level electronic structure theory and experiment,
thus showing promise as a basis for `next-generation' force field development.
%% Based on this accuracy, and using \co as an example, we show preliminary
%% results as to how \mastiff can be used as a basis for next-generation ab
%% initio force field development.
%% \mastiff is transferable and requires minimal additional
%% parameterization, and thus is
%% well suited for future use in next-generation ab initio force field development.

\end{abstract}
%%%%%%%%%%%%%%%%%%%%%%%%%%%%%%%%% Abstract %%%%%%%%%%%%%%%%%%%%%%%%%%%%%%%%%%%%%%%%

%%%%%%%%%%%%%%%%%%%%%%%%%%%%%%%%% Introduction %%%%%%%%%%%%%%%%%%%%%%%%%%%%%%%%%%%%
\begin{section}{Introduction}
\label{sec:intro}

%auto-ignore
Classical molecular simulation is a standard tool for
interpreting and predicting the chemistry of an incredible host of systems ranging
from simple liquids to complex materials and biomolecules. Such simulations
always require, as input, a mathematical description of the
system's potential energy surface (PES). In principle, the
PES for most chemical systems can accurately be determined 
from one of several high-level electronic structure methods;
\cite{Rezac2016,Chalasinski2000,Jan2016}
nevertheless, these calculations are currently too expensive to use
in simulations of large systems and/or long timescales. 
\cite{Hassanali2014}
Consequently, most routine
molecular simulations are performed with the aid of force fields:
computationally-inexpensive, parameterized mathematical expressions that
approximate the exact PES. Because the accuracy and predictive capabilities of
molecular simulations are directly tied to the underlying force
field, a central challenge
is the development of highly accurate force fields. 
In contrast to the development of empirical force fields, where the typical
emphasis is on generation of \emph{effective} potentials yielding bulk
properties, for ab initio force fields,
%Specifically for ab initio force fields, 
this accuracy is principally defined by a force field's fidelity to the underlying exact PES.

As of now, several common
shortcomings inhibit the accuracy and predictive capabilities of
standard ab initio force fields, and these limitations must be systematically addressed
in order to generate improved, `next-generation' force fields.\cite{Zgarbova2010} One important
shortcoming, and the focus of this work, is the so-called
`sum-of-spheres', or `isotropic atom-atom' approximation,\cite{Stone1988} in
which it is presumed that the
non-bonding interactions between molecules can be treated as a superposition
of interactions between \emph{spherically-symmetric} atoms.
(Note that this sum-of-spheres approximation is
distinct from the commonly-used pairwise additive
approximation employed in force fields lacking explicit
polarization;\cite{Zgarbova2010} 
challenges associated with this latter approximation are reviewed elsewhere.
\cite{Lopes2009,Panagiotopoulos2000,Wang2011a,Demerdash2014,Lopes2015,
Stone2007,Price2010a,Welch2008})
The sum-of-spheres approximation thus assumes
that the 
pair potential, $E_2^{ij}$, between two atoms-in-molecules $i$ and $j$
(which
formally depends both on their interatomic distance, $r_{ij}$, and relative
orientation,
$\Omega_{ij}$),
can be modeled as
\begin{align}
\label{eq:sos_approx}
E^{ij}_2(r_{ij},\Omega_{ij}) \approx f(r_{ij}) \equiv V_2(r_{ij}),
\end{align}
where $f(r_{ij})$ is an arbitrary, distance-dependent function that defines the pairwise
interaction.
Here and throughout, we use $E$ to denote an exact PES, and $V$ to denote the
corresponding model/force field energy. 
With some exceptions (vida infra), nearly all standard intermolecular force
fields ---
ranging from the popular ``Lennard-Jones plus point charges'' model to more
complex and/or polarizable functional forms\cite{Schmidt2015}
---
explicitly make use of the isotropic atom-atom approximation.

Notwithstanding its popularity, there is good experimental and
theoretical evidence to suggest that the sum-of-spheres approximation does not
hold in practice.\cite{stone2013theory,Stone1988,Price2000} 
Importantly, and as we argue in
\secref{sec:results}, force fields which account for anisotropic long-range (multipolar) electrostatics, but
otherwise employ the sum-of-spheres approximation, are an improved but
\emph{still incomplete} model for
describing the atomic-level anisotropy of intermolecular interactions.
Experimentally, it has long been known that 
atom-in-molecule charge densities, as determined from x-ray diffraction, can exhibit significant
non-spherical features, such as with lone pair or $\pi$ electron
densities.\cite{Coppens1979} Furthermore, statistical analyses of the
Cambridge Structural Database 
% cite this? is it the cambridge structural database or crystal structure
% database?
have shown that the the van der Waals radii of atoms-in-molecules (as measured
from interatomic closest contact distances)
are not isotropically
distributed, but rather show strong orientation dependencies, particularly for
halogens and other heteroatoms.
\cite{Bondi1964,Nyburg1985,Batsanov2001,Auffinger2004,Lommerse1996,Eramian2013} 
These experimental studies are corroborated by a significant body of
theoretical research on both the anisotropy of the atomic van der Waals radii
as well as the non-spherical features of the atomic charge densities
themselves,
\cite{Wheatley2012,Kramer2014,Lommerse1996, Badenhoop1997a,Kim2014b,Bankiewicz2012}
overall suggesting
that the sum-of-spheres approximation is an insufficiently
flexible model for the subset of intermolecular interactions that arise from
atomically non-spherical charge
densities. 
The breakdown of the sum-of-spheres approximation may be
particularly problematic for ab initio force field development, since any
anisotropy cannot easily be accounted for in an average manner via empirical
parameterization,
and may help explain known difficulties in generating accurate
atom-atom force fields for such important chemical interactions as 
hydrogen bonding,\cite{Cisneros2016a}
$\pi$-interactions,\cite{Chessari2002,Sponer2013,Sherrill2009}
and
$\sigma$-bonding\cite{Bartocci2015,Rendine2011,Politzer2008}
(see \citen{Cardamone2014} and references therein).

Motivated by these observations,
a small but important  body of work has been
devoted to addressing the limitations of the isotropic atom-atom
model in the context of `next-generation' force field development.
As will be discussed in
detail below (see \secref{sec:prior_work}), the general conclusion from these
studies is that many components of intermolecular interactions
(specifically electrostatics, exchange-repulsion, induction, and
dispersion)
can be more accurately modeled by functional forms that go beyond the
sum-of-spheres approximation.
\cite{Price2000,Hagler2015,Ren2003}
While few intermolecular potentials (and virtually no standard force fields
amenable to routine molecular simulation) explicitly account for atomic-level anisotropy
for all aspects of intermolecular interactions, several recent standard force fields
have incorporated atomic-level anisotropy into their description of 
long-range electrostatics.\cite{Cardamone2014} Some of these potentials
(notably AMOEBA\cite{Ponder2010,Ren2003,Shi2013} and some water
potentials\cite{Cisneros2016a,Cardamone2014}) 
are already employed in large-scale molecular simulation, often with very
encouraging success.\cite{Cardamone2014}
Furthermore, others have shown that anisotropic potentials (some of which additionally
model the anisotropy of exchange-repulsion
and/or dispersion) lead to significant improvements in predicting 
molecular crystal structures.
\cite{Cardamone2014,Price2010a,Day1999,Day2003,Price2008,Misquitta2016,Misquitta2008a}
These and other results strongly 
suggest that a complete incorporation of atomic anisotropy 
will lead to increasingly accurate and
predictive molecular simulations in a wider variety of chemical interactions.
\cite{Hagler2015}

Given the
importance of atomic-level anisotropy in defining intermolecular
interactions,
and the critical role that computationally-affordable standard force
fields play in enabling molecular simulation, our present goal is to
develop a general methodology for standard
force field development that can comprehensively account for atomic-level
anisotropy in all
components of intermolecular interactions \emph{and} that can be routinely
employed in large-scale molecular simulation. 
Furthermore, 
our aim is to develop a first-principles-based model that is as accurate and transferable as possible, all while
maintaining a simple, computationally-tractable functional form that allows
for robust parameterization and avoids over-/under-fitting.
Thus, building on prior work (both our own
\cite{VanVleet2016,Schmidt2015,Misquitta2014,Stone2007} 
and from other groups
\cite{Price2000}),
we present here a general ansatz for anisotropic force field development
that, at minimal computational overhead, and only where necessary, incorporates atomic-level anisotropy
into all aspects of intermolecular interactions (electrostatics,
exchange, induction, and dispersion), 
not only in the asymptotic limit of large intermolecular separations, 
but also in the region of non-negligible electron density
overlap.
After motivating and establishing the
functional forms used in our anisotropic force fields, we next demonstrate,
using a large library of dimer interactions between organic
molecules, the accuracy and transferability of these new force
fields with respect to the reproduction of high-quality ab initio potential
energy surfaces. 
Lastly, and using \co as a case study, we offer an example as to
how these new, `atomically-anisotropic' models for dimer interactions
can be used to enable highly accurate
simulations of bulk properties.
The theory
and results presented in this manuscript should be of general utility in
improving the
accuracy of (specifically ab initio generated) force fields, including those
amenable to large-scale molecular dynamics simulations.

\end{section}
%%%%%%%%%%%%%%%%%%%%%%%%%%%%%%%%% Introduction %%%%%%%%%%%%%%%%%%%%%%%%%%%%%%%%%%%%

%%%%%%%%%%%%%%%%%%%%%%%%%%%%%%%%% Prior Work %%%%%%%%%%%%%%%%%%%%%%%%%%%%%%%%%%%%%%
\begin{section}{Background}
\label{sec:prior_work}

Before presenting our development methodology for atomically-anisotropic
potentials, we provide an overview of prior approaches that go beyond
the sum-of-spheres approximation.
Throughout this discussion, we employ the fairly standard\cite{Phipps2015}
decomposition of interaction energies into physically-meaningful components of electrostatics,
exchange-repulsion, induction (which includes both polarization and
charge-transfer), and dispersion. Many studies on atomically-anisotropic force
field development have focused on 
incorporating anisotropy on a component-by-component basis, and so for clarity we discuss
anisotropy for each energy component individually. 
As in prior work,\cite{VanVleet2016} we find it useful to separate 
the so-called `long-range'/asymptotic effects (multipolar electrostatics,
polarization,
and dispersion)
from those `short-range' effects that arise only at
smaller intermolecular separations due to the non-negligible overlap of
monomer electron densities (e.g. charge penetration and exchange-repulsion).

\begin{subsection}{Prior Models for Long-Range Interactions}

The importance of atomic-level anisotropy in long-range interactions,
particularly as it pertains to electrostatics, is quite well known. A number
of groups have found that using atomic multipoles (rather than simple point
charges) greatly improves both the electrostatic potential\cite{Williams1988,Kramer2014} and 
the resulting electrostatic interaction energies.
\cite{Cardamone2014,Ren2003,Shi2013,Demerdash2014,Chaudret2014a,Giese2013,Cisneros2006,Elking2010}
Though not without additional computational cost, atomic multipoles are now
routinely employed in a number of popular force fields.
\cite{Ren2003,Shi2013,Cisneros2016a} 
% Cite other multipolar force fields?
As an alternate and often more computationally-affordable approach, others have
used off-atom point charges to effectively account for anisotropic charge
densities.
\cite{Dixon1997,Harder2006,Rendine2011,Chaudret2013}
In line with chemical intuition, 
improvements from use of atomic multipoles/off-site charges are typically most
significant when describing the
electric fields generated by heteroatoms and carbons in multiple bonding environments.
\cite{Mu2014,Wikfeldt2013}

The induction and dispersion energies have also been shown to exhibit
anisotropies that go beyond the sum-of-spheres model. For instance, it has
been suggested that anisotropic polarizabilities (which affect both polarization
and dispersion) are required to avoid an artificial over-stabilization of base
stacking energies in
biomolecules.\cite{Sponer2013} 
In order to more accurately treat polarization, several molecular mechanics potentials have
made use of either off-site\cite{Piquemal2007} or explicitly anisotropic
polarizabilities.\cite{Harder2006,Loboda2016}. 
Similarly, the importance of anisotropic dispersion interactions has also been
established,
\cite{Misquitta2008,Langhoff1971,Williams2003,Stone2007,Krishtal2011}
particularly for $\pi$-stacking interactions,\cite{Sponer2013,Zgarbova2010}
and select potentials have incorporated directional dependence into the functional
form for dispersion by expanding the dispersion coefficients in terms of 
\sfunc (see \appendixref{sec:appendix}) or, more approximately, 
spherical harmonics.\cite{Williams2003,Stone1984,Stone1978,Misquitta2008,Misquitta2016} 
% Cite s-functions

\end{subsection}
\begin{subsection}{Prior Models for Short-Range Interactions}

At closer intermolecular separations, where overlapping electron densities between
monomers leads to exchange-repulsion and charge-penetration effects,
anisotropy can also be
important. Exchange-repulsion has known orientation
dependencies which can play a quantitative role in 
halogen bonding\cite{Bartocci2015,Stone2013} and other chemical interactions, and many
%TODO: look up importance of exchange-repulsion for other types of
%interactions
authors have developed models for describing the
anisotropy of exchange-repulsion.
Some potentials (albeit not those 
amenable to large-scale molecular simulation) employ numerically computed overlap
integrals in conjunction with 
the density-overlap model popularized by
\citeauthor{Wheatley1990}\cite{Wheatley1990,Kita1976a,Kim1981,Nyeland1986,Ihm1990}
to quantify anisotropic exchange-repulsion, charge transfer, and/or charge
penetration interactions.
\cite{Duke2014a,Cisneros2006,Elking2010,Chaudret2014a,Gavezzotti2003,Torheyden2006}
Taking a more analytical approach, many other potentials have extended the Born--Mayer functional
form\cite{Born1932} to allow for orientation-dependent pre-factors,
\cite{Stone2007,Mitchell2001,Price2000,Stone1988,Day2003,Torheyden2006,Totton2010,Misquitta2016,Price2010a} 
and model short-range effects using an anisotropic functional form originally
proposed by \citet{Stone1988}:
\begin{align}
\vrep_{ij} = G\exp[-\alpha_{ij}(R_{ij} - \rho_{ij}(\Omega_{ij}))].
\end{align}
Here $G$ is not a parameter, but rather an energy unit,\cite{stone2013theory}
$\Omega_{ij}$ describes a relative orientation,
and $\alpha$ and $\rho$ represent, respectively, the hardness and shape of the
pair potential. In principle, one might also allow $\alpha$ to have orientation
dependence; however, this seems unnecessary in practice.\cite{stone2013theory} 
Similar to treatments of
anisotropic electrostatics, the
orientation dependence of $\rho_{ij}$ is typically expressed in terms of spherical harmonics
and/or $\bar{S}$-functions.\cite{stone2013theory}

Finally, we note that, aside from exchange-repulsion, 
we are aware of relatively little research on the development of simple
analytical expressions for the anisotropy of other overlap effects, such as electrostatic/inductive
charge penetration, charge-transfer, or short-range dispersion. 

\end{subsection}

\end{section}
%%%%%%%%%%%%%%%%%%%%%%%%%%%%%%%%% Prior Work %%%%%%%%%%%%%%%%%%%%%%%%%%%%%%%%%%%%%%

%%%%%%%%%%%%%%%%%%%%%%%%%%%%%%%%%%% Theory %%%%%%%%%%%%%%%%%%%%%%%%%%%%%%%%%%%%%%%%
\begin{section}{Theory and Motivation}
\label{sec:theory}

Building on this prior work,
we
now outline a methodology whereby atomic-level anisotropy can be incorporated
into standard force fields amenable to large-scale molecular
simulation. 
%% Before continuing, we must note that, especially given the diversity of prior approaches
%% used to treat atomic-level anisotropy, there is likely no one correct answer
%% as to how force fields should be extended to move beyond the sum-of-spheres
%% approximation. Indeed, we do not wish to provide the reader with the false
%% impression that our ansatz is the best, or only, way to model anisotropy.
%% Rather, 
In particular, we present a general methodology that optimally incorporates
atomically-anisotropic effects subject to the following 
goals:

%TODO: Find a good place to include a discussion of ab-inito vs. empirical
%force field development

\begin{enumerate}
\item \textbf{Chemical accuracy with respect to ab initio benchmarks:} For systems that can be directly parameterized against
high quality ab initio PES, the force field should exhibit chemical
accuracy (average errors smaller than 1\kjmol)
with respect to the ab initio benchmark; furthermore, any errors in the force
field should be random rather than systematic
%% \item \textbf{Chemical accuracy with respect to experiment:} Force fields for a given
%% compound should achieve chemical accuracy with respect to experimental
%% properties across all known phases (solid, liquid, supercritical, etc.) 
\item \textbf{Transferability across chemical environments:} Given force fields for two different pure systems, we
should be able to accurately calculate (via simple
combination rules and without additional
parameterization) the PES of any system that
is a mixture of the pure systems
%% \item \textbf{Universality:} There should be one universally-applicable (and
%% ideally automatable) methodology for force field
%% development that allows us to model (without over-reliance on chemical intuition or ad-hoc
%% inclusion of molecule-specific terms, parameters, or interaction sites) all
%% manner of intermolecular interactions; in other words, the same functinal
%% forms and parameterization scheme should be equally applicable to descriptions
%% of lone pairs, $\pi$ electrons, $\sigma$ holes, etc.
%% \item \textbf{Robustness:} The chosen force field development methodology should not be
%% overly sensitive to the details of parameterization, and the chosen
%% parameterization scheme should not result in either under- or over-fitting of
%% the potential
\item \textbf{Simplicity:} The force field should be restricted to functional forms
that are already compatible with, or could be easily implemented in, existing
molecular simulation packages
\item \textbf{Computational tractability:} The force field should impose
minimal additional computational expense
relative to existing polarizable multipolar force fields\cite{Shi2013}
\end{enumerate}

Given these goals, we now outline a detailed methodology for incorporating
atomic-level anisotropy into each component (electrostatic,
exchange-repulsion, induction, and dispersion) of intermolecular interactions.

\begin{subsection}{Anisotropic Models for Short-Range Interactions}

% \begin{subsubsection}{A Note on Coordinate Systems}
% \end{subsubsection}

\begin{subsubsection}{Exchange-Repulsion}
\label{sec:exchange_theory}

We begin by considering the exchange-repulsion, $\erep_{ij}$, that arises from the overlap of
electron densities from two non-spherical atoms-in-molecules, $i$ and $j$. Here and
throughout, we closely follow the notation and theory used by
\citeauthor{stone2013theory}.\cite{stone2013theory}
Without loss of
generality, we can express the exchange repulsion between these two atoms as a
function of their interatomic distance, $r_{ij}$, and relative
orientation, $\Omega_{ij}$. Furthermore, we can describe this relative
orientation by assigning local coordinate axes to each $i$ and $j$, such that
the exchange energy is given by
\begin{align}
\label{eq:general_anisotropic_repulsion}
\erep_{ij}(r_{ij},\Omega_{ij}) \equiv
\erep_{ij}(r_{ij},\theta_i,\phi_i,\theta_j,\phi_j),
\end{align}
where $\theta_i$ and $\phi_i$ are the polar coordinates, expressed in the local
coordinate system of atom $i$, that describe the position of atom $j$.
Correspondingly, $\theta_j$ and $\phi_j$ define the position of $i$ in terms
of the local coordinate system of $j$. In principle the choice of these local coordinate
frames is arbitrary. However, for the models introduced below,
parameterization can be dramatically simplified by exploiting the
local symmetry of an atom in its molecular environment and aligning the local
coordinate frame with the principal axis of this local
symmetry.\cite{stone2013theory} Some examples of these local axes are shown in
\figref{fig:local_axis}.

          \begin{figure}
          \includegraphics[width=0.9\textwidth]{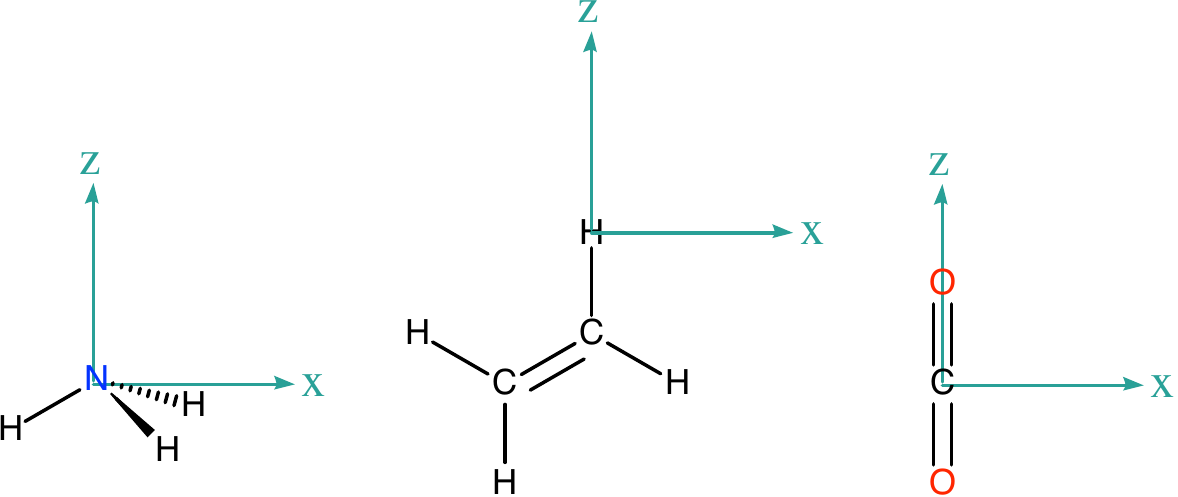}  
          \caption{Local axis system, shown for select atoms in molecules.}
          \label{fig:local_axis}
          \end{figure}

We next make an ansatz that \eqref{eq:general_anisotropic_repulsion} is
separable into radial- and angular-dependent contributions,
\begin{align}
\label{eq:separable_anisotropic_repulsion}
%\erep_{ij}(r_{ij},\theta_i,\phi_i,\theta_j,\phi_j) \approx f(r_{ij})g(\theta_i,\phi_i,\theta_j,\phi_j),
\erep_{ij}(r_{ij},\theta_i,\phi_i,\theta_j,\phi_j) \approx 
\vrep_{ij}(r_{ij},\theta_i,\phi_i,\theta_j,\phi_j) = \fij \gij
\end{align}
thus subdividing the problem of finding a general functional form for $\erep_{ij}$ into
two more tractable tasks. First, we must find an ideal sum-of-spheres model to
describe the radial (isotropic) dependence of the force field, and second, we must find
a way to model the orientation dependence as a multiplicative
pre-factor to \fij.

% TODO: Include a discussion of additional forms for \fij that might serve as
% good functional forms upon which gij can be added?
Given that the only requirement for \fij is that it be isotropic, how should a
suitable model for \fij be chosen?
Indeed, all standard isotropic force fields are of this general form, and thus
might serve as a suitable starting point for anisotropic force field development.
For reasons discussed below, in this work we employ 
a simple and accurate model\cite{VanVleet2016} for \fij that can be
derived from first-principles. In particular, we employ the overlap model
\cite{Kim1981,Nyeland1986,Ihm1990,Stone2007,Wheatley1990,Mitchell2000,Soderhjelm2006,Day2003}
to approximate $\erep_{ij}$ as proportional to
the overlap %, $S_{\rho}^{ij}$,
between spherically-symmetric atom-in-molecule (AIM) electron densities, each with density
\begin{align}
\rho_i(r) = D_i \exp^{-B_i r},
\end{align}
where $D_i$ and $B_i$ are both atom type-specific constants that can be parameterized from
molecular electron densities and that represent, respectively,
the shape and hardness of the AIM density. Using this approximation to the
overlap model,
the exchange energy between two atoms is then modeled by
\begin{align}
\begin{split}
\label{eq:fij}
\erep_{ij} \approx \vrep_{ij} &\propto S_{\rho}^{ij} \\
&\approx \Aex{ij} \left( \frac{ (\B\R)^2}{3} + \B\R + 1 \right) \exp(-\B\R) 
\end{split}
\intertext{with combining rules} 
\begin{split}
\label{eq:aij}
\Aex{ij} 
&\equiv \Aex{i}\Aex{j},\\
\B &\equiv \sqrt{B_iB_j}, \\
\end{split}
\end{align}
and where $S_{\rho}^{ij}$ is the electron density overlap between atoms and \A is
a fitted proportionality constant.

Here and throughout we use \eqref{eq:fij}, referred to 
as the Slater-ISA formalism,\cite{VanVleet2016} as our
model for \fij. This choice is primarily justified by the
accuracy of the Slater-ISA formalism as compared to other
sum-of-spheres models for repulsion.\cite{VanVleet2016} 
Furthermore, and especially for simple test cases where one
might expect the sum-of-spheres approximation to hold (e.g. argon, methane, or ethane), we have shown 
that the \isaffold
correctly models intermolecular potential energy surfaces for a sizable library of intermolecular
interactions over the asymptotic, attractive, and repulsive regions of
the PES.\cite{VanVleet2016}

There is also solid theoretical motivation to utilize Slater-ISA as a model
for \fij.
Specifically, the AIM densities used to parameterize \isaffold are
partitioned using an iterated stockholder atoms (ISA) procedure,
and the resulting density profiles are guaranteed to be maximally spherical.
\cite{Misquitta2014,Lillestolen2008,Lillestolen2009}
%DISCUSS: Alston, to what extent is this true? That is, are the BS-ISA
%densities algorithmically guaranteed to converge to maximally-spherical
%shapes, or does this just empirically seem to be true in practice?
This condition of `maximum sphericity' has two consequences. First, it
suggests that \isaffold should be an optimal, or nearly optimal,
isotropic atom-atom model. In other words, the resulting
model for \fij should completely account for
the radial dependence of the potential, and consequently \gij
will truly represent the orientation dependence, rather than simply over-fitting
residual errors from the radial functional form, in turn retaining high
transferability. 
Second, and relatedly,  having
maximally-spherical ISA densities suggests that anisotropic effects should be a
minimal perturbation to the PES. This means that, to a first-order
approximation, \gij is simply equal to 1. Furthermore, the non-spherical
components of the ISA densities should provide us
with guidance as to which atom types might require anisotropic treatment.

With the functional form for \fij determined, we now describe our
model for \gij. As motivated in \appendixref{sec:appendix}, and under
the ansatz of radial and angular separability,
an approximate, transferable, and orientation-dependent expression for \Aex{i} can be obtained
by expanding \Aex{i} in a
basis of 
renormalized spherical harmonics,
\begin{align}
\label{eq:sph_harm}
C_{lm}(\theta,\phi) = \sqrt{\frac{4\pi}{2l+1}}Y_{lm}(\theta,\phi).
\end{align}
thus yielding
\begin{align}
\label{eq:gij}
\begin{split}
\Aex{i}(\theta_i,\phi_i) &= 
\Aex{i,\text{iso}}\big(1 + \aniso{exch} \big), \\
\aniso{exch} &\equiv \sum\limits_{l>0,k} \aex  C_{lk}(\theta_i,\phi_i)
\end{split}
\end{align}
for \Aex{i} and subsequently
\begin{align}
\label{eq:vex}
\vrep_{ij} &= \Aex{ij}(\Omega_{ij})
        \left( \frac{ (\B\R)^2}{3} + \B\R + 1 \right) \exp(-\B\R)
\end{align}
with
\begin{align}
\Aex{ij}(\Omega_{ij}) &= \Aex{i}(\theta_i,\phi_i)\Aex{j}(\theta_j,\phi_j)
\end{align}
for the exchange-repulsion potential. Note that, with the exception of the
now orientation-dependent \Aex{i}, the atomically-anisotropic model in \eqref{eq:vex} is identical to
our previously-defined isotropic model (\eqref{eq:fij}).

The \aex are free parameters which must be fit to ab initio
data. 
Still, we and others have found the expansion in \eqref{eq:gij} to be very quickly convergent,
\cite{Stone2007,Mitchell2001,Price2000,Stone1988,Day2003,Torheyden2006,Totton2010,Misquitta2016,Price2010a}
 especially given a proper choice of local coordinate
system that eliminates many expansion terms via symmetry. In practice, only
symmetry-allowed terms up
to $l=2$ seem to be required for heteroatoms, carbons in multiple bonding
environments, and select hydrogens (see
equations in \secref{sec:results}). Most other atom types require no anisotropic
parameters whatsoever, 
and isotropic atom types can be easily modeled within
this formalism simply by setting $\aniso{} = 0$.

\end{subsubsection}
\begin{subsubsection}{Other Short-Range Effects}

As in prior work,\cite{VanVleet2016} we have found that other short-range
effects, including charge
penetration and short-range induction, can be modeled as proportional to
exchange-repulsion. We take the same approach here, and
the functional form for these two short-range effects is given by
\eqref{eq:vex}, with `exch' superscripts replaced by the appropriate short-range
energy term (see \secref{sec:methods}).

At shorter intermolecular separations, we must also damp some of the
functional forms developed for long-range interactions (vida infra) so as to
account for charge penetration effects and avoid unphysical divergences.  For
induction, we use the same isotropic damping function as in the AMOEBA force
field.\cite{Shi2013}
To model the dispersion energies at short-range, we
damp each of the individual $C_{n}$ dispersion coefficients (see
\cref{sec:dispersion} and \cref{eq:ff_form}) with
the Tang-Toennies\cite{Tang1984,Tang1992} damping function,
\begin{align}
\begin{split}
f_{n}(x) &= 1 - e^{-x} \sum \limits_{k=0}^{n} \frac{(x)^k}{k!} \\
x &= -\frac{d}{d r_{ij} }\left[\ln \vrep_{ij} \right] \ r_{ij}.
\end{split}
\end{align}
By substituting in our expression for $\vrep_{ij}$ from \cref{eq:vex}, we obtain
\begin{align}
x = B_{ij}r_{ij} - \frac{2 B_{ij}^2 r_{ij} + 3 B_{ij} }
{ B_{ij}^2 r_{ij}^2 + 3 B_{ij} r_{ij} + 3} r_{ij} 
\end{align}
for our anisotropic model,
which is identical to the expression derived for isotropic systems in prior work.\cite{VanVleet2016}

\end{subsubsection}
\end{subsection}

\begin{subsection}{Anisotropic Models for Long-Range Interactions}
\label{sec:lr_theory}

\begin{subsubsection}{Electrostatics}

In the present work, 
we describe the asymptotic electrostatics via a distributed multipole expansion,
\cite{stone2013theory,Stone2007}
\begin{align}
\label{eq:electrostatic_t}
V^{\text{multipole}}_{ij} = \vmultipole
\end{align}
with multipolar interaction tensor $T$ and parameterized moments $Q$ for all
multipole moments $tu$ up to rank 2.
However, for increased computational efficiency, off-site point charge models
could also be utilized.
\cite{Cardamone2014}

\end{subsubsection}
\begin{subsubsection}{Induction}
\label{sec:polarization}

Just as with electrostatics, long-range induction should properly be described
by a distributed multipole expansion of interacting atomic
polarizabilities.\cite{Stone2007,Misquitta2016} Indeed, it has been shown
that inclusion of higher-order and/or anisotropic polarizabilities greatly
reduces errors in the two-body induction potential relative to commonly-used isotropic
dipole polarizability models.\cite{Misquitta2008b,Holt2008,Holt2010,Schmidt2015,Shi2013}
Because the model for the two-body induction also determines the many-body
polarization energy, the proper treatment of induced multipoles becomes especially
important in condensed phase simulation.\cite{stone2013theory,Schmidt2015,Shi2013}

Owing to the increased computational cost of these higher-order and anisotropic polarizability
models, and because such functional forms are 
not (yet) implemented in OpenMM (the molecular simulation package used in
this work), we
currently
neglect
both higher-order and anisotropic contributions to the long-range induction. 
%% Thus, in the present work, induction is described by the isotropic induced dipole formalism,
%% %
%% \begin{align}
%% \vdrude = -\sum\limits_{i=1}^{N} \vec{\mu}_i\cdot
%% \end{align}
%% %
%% 
As
we shall show, however, errors in the induction potential limit the overall accuracy of
our force fields for extremely polar molecules (notably water), and
future work will likely require improved models for long-range induction. 

\end{subsubsection}
\begin{subsubsection}{Dispersion}
\label{sec:dispersion}

Past research\cite{stone2013theory} has motivated an anisotropic atom-atom model for dispersion 
of the form
%
% TODO: Check the limits on this equation
\begin{align}
\label{eq:stone_disp}
\vdisp_{ij} = - \sum\limits_{n=6} \frac{C_{ij,n}(\Omega_{ij})}{\R^n}.
\end{align}
Note that, in this equation, both odd and even powers of $r$ are allowed in the
dispersion expansion, where all coefficients associated with odd powers are
non-zero only for anisotropic charge distributions.
In order to make this model both computationally efficient and maximally compatible with our previous isotropic
model for dispersion, 
%% and in a manner analogous to our treatment of
%% short-range overlap effects, 
we choose (as an ansatz) to model the dispersion anisotropy as an
orientation-dependent \emph{prefactor} that affects all isotropic $C_6 - C_{12}$ dispersion
coefficients equally:
\begin{align}
\label{eq:mastiff_disp}
\vdisp_{ij} &= - \Adisp{i}\Adisp{j}\sum\limits_{n=3}^{6} \frac{C_{ij,2n}}{\R^{2n}} 
\intertext{with}
\label{eq:mastiff_disp2}
\Adisp{i} &= 1 + \aniso{disp}
\end{align}
and \aniso{disp} as in \eqref{eq:gij}.
%% Because $C_{00}(\theta,\phi) = 1$, the anisotropy expansion proposed above has
%% no effect on the average magnitude of the various dispersion coefficients.
Once again, \eqref{eq:mastiff_disp} reduces to the isotropic case by setting
$\aniso{disp} = 0$.
We note that, though the functional form in \eqref{eq:mastiff_disp} bears many similarities to
\eqref{eq:stone_disp}, (unphysically) no odd powers of $r$ occur
in our proposed model for dispersion. 
Furthermore, 
the model utilizes the same anisotropic expansion for each dispersion
coefficient.
Nonetheless, we will show in \secref{sec:results}
that this simple model yields significant accuracy gains in the dispersion energy
with only minimal additional parameterization and expense.

\end{subsubsection}

\end{subsection}

\end{section}
%%%%%%%%%%%%%%%%%%%%%%%%%%%%%%%%%%% Theory %%%%%%%%%%%%%%%%%%%%%%%%%%%%%%%%%%%%%%%%

%%%%%%%%%%%%%%%%%%%%%%%%%%%% Computational Details %%%%%%%%%%%%%%%%%%%%%%%%%%%%%%%%
\begin{section}{Technical Details}
\label{sec:methods}

\begin{subsection}{The 91 Dimer Test Set}

Our benchmarking procedures
are the same as in prior work,\cite{VanVleet2016}
and we briefly summarize the relevant technical details. A full discussion of
results and example calculations are presented in \secref{sec:results}.

We have previously developed a large library of benchmark interaction energies
involving the following 13 atomic and small organic species: acetone, argon, ammonia, carbon dioxide,
chloromethane, dimethyl ether, ethane, ethanol, ethene, methane, methanol,
methyl amine, and water. (As in prior work, these molecules were chosen to be broadly
representative of various functional groups in organic chemistry; studies on
larger and/or flexible molecules are outside of the scope of this work, but
will be the subject of future work.) Using these 13 monomers, we have
generated a library of dimer interaction
energies for each of the 91 possible unique dimer combinations (13 homomonomeric, 78
heteromonomeric). For each of these dimer combinations, interaction energies were
computed at a DFT-SAPT
\cite{Misquitta2002,Misquitta2003,Misquitta2005,Heßelmann2005a,Podeszwa2006a,Heßelmann2002,Heßelmann2003,Heßelmann2002a,Jansen2001}
level of theory for 1000 quasi-randomly chosen dimer configurations,
representing 91,000 benchmark interaction energies in total. 
As described below, parameters for a given force field methodology are then
fit on a component-by-component basis to reproduce the benchmark DFT-SAPT energies. 

\end{subsection}
\begin{subsection}{Force Field Fitting}

\begin{subsubsection}{Functional Forms}

We will present three types of force field fitting methodologies in this work,
termed \isoff, \isaff, and \anisoff (also referred to as 
a
\textbf{M}ultipolar, \textbf{A}nisotropic, \textbf{S}later-\textbf{T}ype
\textbf{I}ntermolecular \textbf{F}orce \textbf{F}ield, \mastiff).
The nomenclature of each
name refers to, first, the isotropic/anisotropic treatment of multipolar
electrostatics and, second, the isotropic/anisotropic treatment of dispersion
and short-range effects. 
For \mastiff, dispersion and short-range anisotropies are only included
on heteroatoms, atoms in multiple bonding environments, and associated
hydrogens (see \si).
Note that \isaff is virtually identical to the \isaffold
model developed in our prior work, and that
this \emph{partial} treatment of anisotropy (via multipolar
electrostatic terms) is very similar in spirit
to the popular AMOEBA\cite{Ren2003,Shi2013}
methodology. 

All force fields in this work
use the following general functional form for two-body interactions,
\begin{align}
\label{eq:ff_form}
\vtot^{2b} &= \sum\limits_{ij} \vrep_{ij} + \velst_{ij} + \vind_{ij} + \vdhf_{ij} +
\vdisp_{ij} 
\intertext{where}
\begin{split}
\label{eq:ff_details}
\vrep_{ij} &= \Aex{ij} P(B_{ij}, r_{ij}) \exp(-B_{ij}r_{ij}) \\
\velst_{ij} &= -\Ael{ij} P(B_{ij}, r_{ij}) \exp(-B_{ij}r_{ij}) + \vmultipole
\\
\vind_{ij} &= -\Aind{ij} P(B_{ij}, r_{ij}) \exp(-B_{ij}r_{ij}) + \vdrudeind \\
\vdhf_{ij} &= -\Adhf{ij} P(B_{ij}, r_{ij}) \exp(-B_{ij}r_{ij}) +
\vdrudescf \\
\vdisp_{ij} &= - \Adisp{ij} \sum\limits_{n=3}^{6} f_{2n}(x) \frac{C_{ij,2n}}{r_{ij}^{2n}}
\\
P(B_{ij},r_{ij}) &= \frac13 (B_{ij} r_{ij})^2 + B_{ij} r_{ij} + 1 \\
\A &= A_iA_j \\
\B &= \sqrt{B_iB_j} \\
\C &= \sqrt{C_{i,2n}C_{j,2n}} \\
f_{2n}(x) &= 1 - e^{-x} \sum \limits_{k=0}^{2n} \frac{(x)^k}{k!} \\
x &= B_{ij}r_{ij} - \frac{2 B_{ij}^2 r_{ij} + 3 B_{ij} }
{ B_{ij}^2 r_{ij}^2 + 3 B_{ij} r_{ij} + 3} r_{ij} ,
\end{split}
\end{align}
$B_i$, $C_i$, and $Q_i$ coefficients are all parameters of the force field
(see \cref{sec:monomer_params} for details), and $T$
is the multipolar interaction tensor given in \appendixref{sec:appendix}. 
For \isoff (the completely isotropic model), the summation in
\vmultipole is truncated to only include point charges, whereas \isaff and \mastiff both
use a multipole expansion up to quadrupoles.

Both \isoff and \isaff treat each $A_i$ as a single
fitting parameter, with the exception that
$\Adisp{i} = 1$. 
By contrast, $A_i$ is modeled in our fully anisotropic model, \mastiff, 
as an orientation-dependent function expressed as an expansion in
terms of spherical harmonics,
\begin{align}
\label{eq:v_aniso}
\begin{split}
A_{i}(\theta_i,\phi_i) &= 
A_{i,\text{iso}}\big(1 + \aniso{} \big), \\
\aniso{} &\equiv \sum\limits_{l>0,k} a_{i,lk}  C_{lk}(\theta_i,\phi_i),
\end{split}
\end{align}
where $A_{i,\text{iso}}$ and $a_{i,lk}$ are fitted parameters. As with the
previous two force fields, 
$A_{i,\text{iso}}^{\text{disp}} = 1$ for \mastiff. For isotropic atom types
in \mastiff (listed in the \si), $\aniso{} = 0$, such that the functional form for
\emph{isotropic} atomtypes is identical between \mastiff and \isaff, and only
the functional form for \emph{anisotropic} atom types differ between force
fields.  Note, however, that the numerical values for $A_{i,\text{iso}}$
in \mastiff can differ from that of the $A_i$ parameters used in the other models.

As in \citen{McDaniel2013}, and for the purposes of force field fitting, the polarization energy,
$\vdrude = \vdrudeind + \vdrudescf$,
is
calculated using using a Drude oscillator model. 
As a difference from prior work, here the Thole-damping function 
follows the same functional form as in the AMOEBA model,\cite{Ren2003} 
\begin{align}
\rho = \frac{3a}{4\pi} \exp(-au^3),
\end{align}
where $a=0.39$ is a damping parameter, and $u=r_{ij} / (\alpha_i \alpha_j)^{1/6}$ is
an effective damping distance that depends on calculated atomic
polarizabilities (vida infra),
$\alpha_i$. (The choice of damping function was selected
for later compatibility with the OpenMM\cite{Eastman2013} software package; see
\cref{sec:methods_simulations} for details.)
As described fully in
\citen{McDaniel2013},
and for the purpose of logical consistency with the
corresponding SAPT energies (see \cref{sec:sapt}),
during force field fitting
\vdrude is subdivided into 2\textsuperscript{nd} (\vdrudeind) and
higher-order \vdrudescf contributions, and
each contribution to the Drude oscillator energy is then added to either \vind or
\vdhf, respectively.

\end{subsubsection}
\begin{subsubsection}{Benchmark Energies}
\label{sec:sapt}

Because DFT-SAPT provides a physically-meaningful energy decomposition into electrostatic,
exchange-repulsion, induction, and dispersion terms, parameters for each term in
\eqref{eq:ff_form} are directly fit to model the corresponding DFT-SAPT energy
(see \citen{VanVleet2016} and references therein for details on the DFT-SAPT
terminology): 
\begin{align}
\begin{split}
\vrep \approx \erep &\equiv E^{(1)}_{\text{exch}} \\
\velst \approx \eelst &\equiv E^{(1)}_{\text{pol}} \\
\vind \approx \eind &\equiv E^{(2)}_{\text{ind}} + E^{(2)}_{\text{ind-exch}} \\
\vdhf \approx \edhf &\equiv \delta(\text{HF}) \\
\vdisp \approx \edisp &\equiv E^{(2)}_{\text{disp}} + E^{(2)}_{\text{disp-exch}}.
\end{split}
\end{align}
Fitting parameters on a component-by-component basis helps ensure parameter
transferability and minimizes reliance on error cancellation. Note that no
parameters are fit to reproduce the total energy and that,
because the DFT-SAPT energy decomposition is only calculated to second-order, 
third- and higher-order terms (mostly consisting of higher-order induction)
are estimated by \edhf.

\end{subsubsection}
\begin{subsubsection}{Parameters Calculated from Monomer Properties}
\label{sec:monomer_params}

Of the parameters listed in \eqref{eq:ff_details}, most do not need to be
fit to the DFT-SAPT energies, but can instead be calculated directly on the
basis of monomer electron densities. In particular, all multipolar
coefficients $Q$, polarizabilities $\alpha_i$ (involved in the calculation of \vdrude),
dispersion coefficients $C$, and atom-in-molecule exponents $B^{\text{ISA}}$, are calculated in a manner nearly
identical to \citen{VanVleet2016}. Note that, for our atom-in-molecule exponents, we tested
the effects of treating $B^{\text{ISA}}$ either as a hard- or
soft-constraint in the final force field fit. While the general conclusions from this
study are rather insensitive to this choice of constraint methodology, we have
found that the overall force field quality is somewhat improved by
relaxing the $B^{\text{ISA}}$ coefficients in the presence of a harmonic
penalty function (technical details of which can be
found in the Supporting Information of \citen{VanVleet2016}). The optimized $B$
coefficients in this work are always within 5--10\% of the calculated $B^{\text{ISA}}$
coefficients, demonstrating the good accuracy of the $B^{\text{ISA}}$
calculations themselves.

\end{subsubsection}
\begin{subsubsection}{Parameters Fit to Dimer Properties}
\label{sec:dimer_params}

In addition to the soft-constrained $B$ parameters, all other free parameters
($A$ and $a$ parameters from \eqref{eq:ff_form}
and \eqref{eq:v_aniso}) are fit to reproduce
DFT-SAPT energies from the 91 dimer test set described above. For each dimer
pair, 4-5 separate optimizations (for exchange, electrostatics, induction,
\dhf, and, for \mastiff, dispersion) were carried out to minimize a weighted
least-squares error. with the weighting
function given by a Fermi-Dirac functional
form,
\begin{align}
\label{eq:weighting-function}
w_i = \frac{1}{\exp(\sfrac{-E_i}{5.0 |E_{\text{min}}|}) + 1},
\end{align}
where $E_i$ is the reference energy and 
$E_{\text{min}}$ is 
an estimate of the global minimum well
depth (see \citen{VanVleet2016} for details). 

\end{subsubsection}
\begin{subsubsection}{Local Axis Determination}

Identically to AMOEBA and other force fields that incorporate some degree of
atomic-level anisotropy,\cite{Ren2003,Day2003,Totton2010} we use a z-then-x
convention to describe the relative orientation of atomic species. By design,
the z-axis is chosen to lie parallel to the principal symmetry axis (or
approximate local symmetry axis) of an atom
in its molecular environment, and the xz-plane is similarly chosen to
correspond to a secondary symmetry axis or plane. Based on the assigned symmetry
of the local reference frame, many terms in the
spherical expansion of \eqref{eq:gij} can then be set to zero, minimizing the
number of free parameters that need to be fit to a given atom type. 
%TODO: Add to SI
Representative local reference frames are shown for a few atom types in
\figref{fig:local_axis}, and a complete listing of anisotropic atom types
(along with their respective local reference frames and non-zero spherical
harmonic expansion terms)
are given in the \si.

\end{subsubsection}
\begin{subsubsection}{CCSD(T) Force Fields}

DFT-SAPT is known to systematically underestimate the interaction energies of
hydrogen-bonding compounds, and can also exhibit small but important errors
for dispersion-dominated compounds.\cite{Parker2014} Consequently, for
simulations involving \co, \cl, \nh, and \ho, we tested the effect of
refitting our SAPT-based force
fields to reproduce benchmark supermolecular, counterpoise-corrected CCSD(T)-F12a/aVTZ
calculations on the respective dimers. All calculations were performed using
the Molpro 2012 software.\cite{MOLPRO} As with the DFT-SAPT-based force
fields, all fits were performed on a
component-by-component basis to fit (aside from the dispersion, discussed
below) the corresponding DFT-SAPT energies
as calculated in prior work:\cite{VanVleet2016} 
\begin{align}
\begin{split}
\vrep \approx \erep &\equiv E^{(1)}_{\text{exch}} \\
\velst \approx \eelst &\equiv E^{(1)}_{\text{pol}} \\
\vind \approx \eind &\equiv E^{(2)}_{\text{ind}} + E^{(2)}_{\text{ind-exch}} \\
\vdhf \approx \edhf &\equiv \delta(\text{HF}) \\
\vdisp \approx \edisp &\equiv E^{(2)}_{\text{disp}} +
E^{(2)}_{\text{disp-exch}} + \delta(\text{CC}),
\end{split}
\end{align}
where $\delta(\text{CC}) \equiv E_{\rm{int}}^{\text{CCSD(T)-F12a}} -
E_{\rm{int}}^{\text{DFT-SAPT\phantom{}}}$. 
In the case of dispersion, and so that
the total benchmark energy corresponded to the total CCSD(T)-f12a/aVTZ
interaction energy,
the difference between coupled-cluster and
SAPT energies was added to the SAPT dispersion energy. (This correction scheme
was chosen to account for 
small differences in electron correlation effects between coupled cluster
and
DFT-SAPT.)
The dispersion model \vdisp was then parameterized to reproduce the modified \edisp
energy.

In fitting our CCSD(T)-f12a-based force fields, 
we somewhat relaxed the
constraint that $\Adisp{} = 1$ for all atom types, and instead let $ 0.7 \le
\Adisp{} \le 1.3$. This constraint relaxation led, in some cases, to modest
improvements in the fitted potential.

\end{subsubsection}

\begin{subsubsection}{CO$_2$ 3-body potential}
For modeling bulk \co, we developed a three-body model to
account for three-body dispersion effects. This three-body model is based on
the three-body dispersion Axilrod-Teller-Muto (ATM) type model developed by \citeboth{Oakley2009a}. These
authors fit the ATM term with the constraint that the total molecular $C_9$
coefficient be 1970 a.u. Based on our own calculations using a CCSD/AVTZ
level of theory,\cite{Korona2011} we have obtained an
isotropic molecular $C_9$ coefficient of 2246 a.u.; consequently, a 1.13 universal
scale factor was introduced to the Oakley potential so as to obtain dispersion
energies in line with this new dispersion coefficient.
\end{subsubsection}

\end{subsection}

%% \begin{subsection}{Comparison to Ab Initio Benchmarks}
%% 
%% As in previous work, root-mean-square (\rmse) and mean-signed errors (MSE),
%% both with respect to the DFT-SAPT reference energies,
%% were calculated for each methodology and for each dimer pair. Similarly,
%% `attractive \rmse/MSE' (a\rmse/aMSE) were computed by only considering the
%% subset of dimer configurations with net attractive total energies (as measured
%% by DFT-SAPT). After taking the absolute value of the MSE values, the
%% various error metrics were then averaged in the geometric mean sense to
%% obtain one `characteristic' \rmse or \mse for the entire 91 dimer
%% test set.
%% 
%% \end{subsection}
\begin{subsection}{Simulation Protocols}
\label{sec:methods_simulations}

\begin{subsubsection}{Polarization Models for Simulations}

Though we have used a Drude oscillator model in the past and during force
field development, at present Drude oscillators in the
OpenMM\cite{Eastman2013} software are not compatible with use of higher-order
multipoles. For this reason, here our molecular simulations use an induced
dipole model to describe polarization effects, with functional form identical
to that from the AMOEBA force field.\cite{Ren2003} Numerical differences
between the Drude oscillator and induced dipole models were found to be
negligible. 

\end{subsubsection}
\begin{subsubsection}{2\textsuperscript{nd} Virial Calculations}

Classical second virial coefficients were calculated for \nh, \ho, \co,
and \cl using rigid monomer geometries and following the procedure described in
\citen{McDaniel2013}.

\end{subsubsection}

\begin{subsubsection}{\deltahsub for CO$_2$}

For \co, the molar enthalpy of sublimation was determined according to 
\begin{align}
\begin{split}
\deltahsub &= H_{\text{g}} - H_{\text{crys}}  \\
           &= (U_{\text g} + PV_{\text g}) - (U_{\text{el,crystal,0K}} 
                +\Delta U_{\text{el,crystal,0K}\to T_{\text{sub}}} + PV_{\text{crys}} + E_{\text{vib,crystal}}) \\
           &\approx (RT) - \left(U_{\text{el,crystal,0K}} 
                + \int_{0K}^{T_\text{sub}} C_p dT \quad + E_{\text{vib,crystal}}\right) \\
\end{split}
\end{align}
which assumes ideal gas behavior and $PV_{\text{g}} >> PV_{\text{crys}}$.
For the crystal, an experimental measure of $C_p$ was obtained from \citen{Giauque1937} and numerically
integrated to obtain a value $\Delta U_{\text{el,crystal,0K}\to
T_{\text{sub}}} = 6.70 \kjmol$. Theoretical measures of
$E_{\text{vib,crystal}} \approx 2.24 - 2.6 \kjmol$ were obtained from
(respectively) \citen{Cervinka2017} and \citen{Heit2016a}, 
and
$U_{\text{el,crystal,0K}}$ was determined from the intermolecular force field
using a unit cell geometry taken from experiment.\cite{Simon1980}

\end{subsubsection}
\begin{subsubsection}{Other CO$_2$ Simulations}

To determine the densities and enthalpies of vaporization used in this work,
simulations were run in OpenMM using NPT and NVT ensembles, respectively.
Bulk \co was modeled using 780 rigid \co molecules and periodic boundary conditions. 
Electrostatic interactions were described with the
particle-mesh Ewald (PME) method,
three-body dispersion was treated using a 9\AA{} cutoff,
and the remainder of the potential was computed using a 14\AA{} cutoff
and long-range energy/pressure corrections. A Langevin integrator (with a
friction coefficient of 2.0 ps$^{-1}$) and
Monte Carlo barostat were utilized, when required, for temperature and pressure
coupling.
A cubic box with isotropic coupling was used for NPT
simulations, and a 0.5 fs time step was used for all simulations.
Under these conditions, and using an unoptimized version of OpenMM (see \si
for details), simulations speeds were $\sim$2.5 ns/day (for \mastiff) or
$\sim$3.1 ns/day (for \isaff).
After an equilibration period of at least 100ps, simulation data was collected
for a minimum of 1 ns.
Average densities
were obtained directly from simulation, and 
the molar enthalpy of vaporization for \co was determined from the following
formula:
\begin{align}
\begin{split}
\deltahvap &= H_{\text{g}} - H_{\text{liq}} \\
           &= U_{\text{g}} - U_{\text{liq}} + P(V_{\text{g}} - V_{\text{liq}})
\end{split}
\end{align}
Note that, at the state points studied, the ideal gas approximation is
insufficiently accurate, and thus simulations were run for both the gas and
liquid phases at experimentally-determined
densities and pressures.\cite{Span1996}

\end{subsubsection}
\end{subsection}

\end{section}
%%%%%%%%%%%%%%%%%%%%%%%%%%%% Computational Details %%%%%%%%%%%%%%%%%%%%%%%%%%%%%%%%

%%%%%%%%%%%%%%%%%%%%%%%%%%%%%%%%%% Results %%%%%%%%%%%%%%%%%%%%%%%%%%%%%%%%%%%%%%%%
\begin{section}{Results and Discussion}
\label{sec:results}

\begin{subsection}{Overview}

We now benchmark our
developed anisotropic force field methodology against various
sum-of-spheres approximations. 
As is standard in ab initio force field development, we will principally rely
on the following metric for force field quality:
the accuracy with which a given force field functional form can reproduce
high-quality ab initio benchmark energies. 
Furthermore, our choice of relevant benchmark energies is guided by the 
many-body expansion (MBE),
\cite{Stone2007,Elrodt1997}
whereby the energy of an arbitrary
$N$-particle system
is expressed
as a sum of $n$-body interaction potentials,
\begin{align}
\label{eq:mbe}
E_N(\vec r_1 ,\vec r_2 ,\dots,\vec r_N ) =
    % \sum\limits_{i}^{N} E_1(\vec r_i) +
    \sum\limits_{i < j}^{N} E_2(\vec r_i, \vec r_j) +
    \sum\limits_{i < j < k}^{N} \Delta E_3(\vec r_i, \vec r_j, \vec r_k) +
\dots
\end{align}
$E_2$, the `pair potential', is defined as
the difference in interaction energies between a molecular
dimer and the individual monomers themselves; 
$\Delta E_3$ corresponds to the non-additive contributions (energy
not accounted for in $E_2$) to the interaction energies of trimers, and
higher-order terms in the expansion are defined analogously.  
Aside from many-body polarization, for
which the complete $N$-body effects can readily be
calculated\cite{Stone2007,Rick2002}, the MBE typically converges rapidly,
such that only $E_2$ and occasionally $\Delta E_3$ terms are required to completely
and accurately describe $E_N$.\cite{Stone2007,stone2013theory} 
(Notably, the
combination of $E_2$ and $N$-body polarization often account for
upwards of 90--95\% of the total interaction energy;
\cite{McDaniel2014,stone2013theory} 
as discussed in \secref{sec:co2}, any important contributions from $\Delta E_3$ can be
accounted for separately and systematically using known
methods.\cite{Yu2012b,McDaniel2014})
Thus,
because the accuracy and
predictive power of an ab
initio force field depends substantially on the accuracy with which we can
describe $E_2$, and because 
the
functional forms introduced in \secref{sec:theory} directly affect only this
pairwise-additive portion of the intermolecular potential, 
\emph{we
primarily
concentrate our efforts on assessing force field quality with respect to benchmark
calculations of dimer interaction energies. }

In addition to the above comparisons to ab initio benchmarks, a secondary goal of this work 
is to evaluate the extent to which the force field
methodologies presented here can be used, not only to reproduce ab initio
benchmarks, but also to accurately simulate experimental
properties. Especially for ab initio force fields,
accurate comparisons to experiment depend, not only on the quality of the
two-body force field (as defined above), but also on the accuracy of the
benchmark electronic structure theory, the treatment of many-body and/or
quantum effects, etc. Thus for select systems, we also compare our
force fields to experimental second virial coefficients and bulk properties,
with the goal of offering preliminary insight into how our anisotropic force field
methodology might
be utilized, in conjunction with accurate electronic structure theory and a
proper treatment of many-body effects,
to yield a complete $N$-body force field capable of accurately
simulating experimental properties across a wide range of phase space.

\end{subsection}
\begin{subsection}{Accuracy: Comparison with DFT-SAPT}
\label{sec:accuracy}

We compare between three models in this work (see \cref{sec:methods} for
detailed functional forms): \isoff, which
uses a completely isotropic description of all energy components, \isaff,
which additionally accounts for multipolar electrostatic anisotropy, 
and \mastiff, which incorporates anisotropy into all energy
components of the intermolecular potential. 
For each of the 91 dimer combinations described in \secref{sec:methods}, and
for each model,
parameters were fit to reproduce benchmark \sapt energies calculated for 1000 different
relative orientations of the constituent monomers. 
From these `dimer-specific' fits, and as in
described in our prior work,\cite{VanVleet2016} we then averaged the
root-mean-squared (\rmse) and mean signed errors (\mse) from each of the 91
fits to produce so-called `characteristic \rmse/\mse', metrics
representative of the errors associated with a given force field methodology.
Because the absolute magnitudes of the various energy components
becomes large in the repulsive portion of the potential, these characteristic
errors are dominated by repulsive
configurations. As such,
we have also calculated `attractive
\rmse/\mse' (a\rmse/a\mse), defined as the characteristic errors for the subset
of configurations with total interaction energies $E_{tot} < 0$. 
All computed characteristic \rmse are shown in \figref{fig:rmse}, with \mse
data shown in the \si. Unless otherwise stated, results in this section refer
exclusively to the `Dimer-specific' fits in \figref{fig:rmse}, with an
explanation and full discussion of so-called `Transferable' fits given in
\secref{sec:transferability}.

    %%%%%%%%%%%% Average RMSE %%%%%%%%%%%%%%%
    \begin{figure}
    \includegraphics[width=0.9\textwidth]{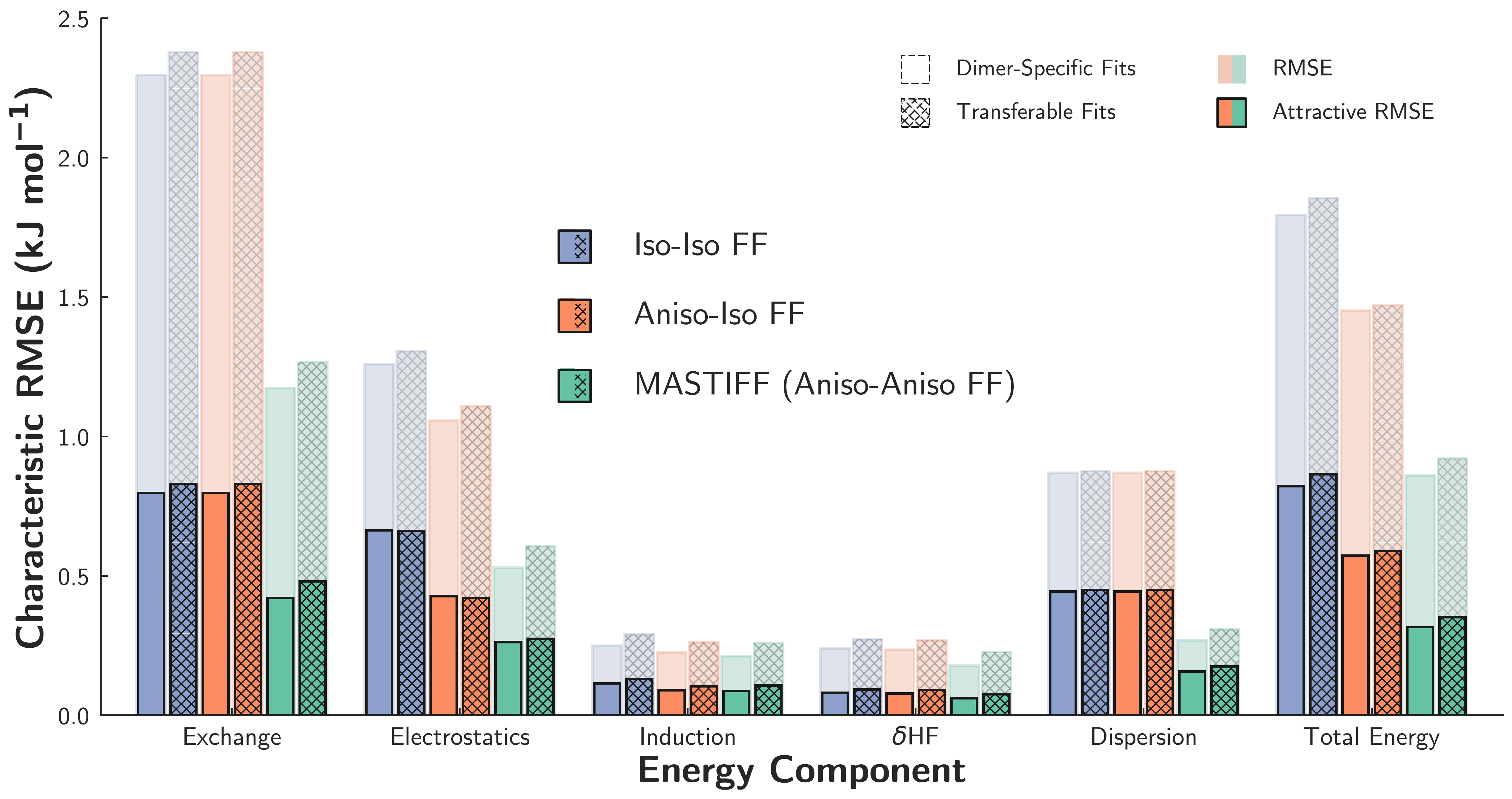}
    \caption{
    Characteristic RMSE (as described in the main text) for the \isoff
(purple), \isaff (orange), and \mastiff (green) over the 91
    dimer test set. The semi-transparent bars represent total RMSE
    for each energy component, while the smaller solid bars represent `Attractive'
    RMSE, in which repulsive points have been excluded. For each force field,
    two types of fits, dimer-specific (solid) and transferable (hashed lines),
    are displayed; see \secref{sec:transferability} for details. Finally, note that, for \isoff and \isaff, only the electrostatic
    and total energy \rmse's differ. 
            }
    \label{fig:rmse}
    \end{figure}
    %%%%%%%%%%%% Average RMSE %%%%%%%%%%%%%%%

Based on the characteristic \rmse shown in \figref{fig:rmse}, both \isaff
and \mastiff offer substantial improvements over the completely isotropic model
\isoff. Though unsurprising, given the well-studied importance of higher-order
electrostatic multipole moments, \isaff shows reduced \rmse/a\rmse that are (depending on the exact
error metric used)
roughly 30\% smaller than \isoff. 
Both \rmse and a\rmse measures showing similar gains in accuracy,
indicating that inclusion of higher-order multipoles (henceforth
`multipolar electrostatic anisotropy') is important in both
attractive and repulsive regions of the potential. Crucially, inclusion of
additional `short-range anisotropies' (anisotropic
interactions arising from overlap of monomer electron densities, namely
exchange-repulsion and electrostatic/inductive charge penetration) and
long-range `dispersion anisotropy' yields a \emph{further} 40\% reduction in
\rmse/a\rmse for \mastiff as compared to the \isaff.
This latter
result is highly important, as it suggests that, for the generation of highly
accurate ab initio potentials, the combination of short-range
and dispersion anisotropies 
are comparable in importance to
multipolar electrostatic anisotropy. Indeed, this substantial increase in
force field accuracy, arising from a \emph{comprehensive} treatment of anisotropic
effects, 
is one of the most important findings in the present work.
In summary, and encouragingly, the combination of multipolar electrostatic, short-range, and
dispersion anisotropies result in an overall 60\% reduction in \rmse/a\rmse when
comparing \isoff to \mastiff.

\figref{fig:rmse} also displays characteristic \rmse/a\rmse for each 
component of the force field, allowing us to account for the influence of
anisotropy on a term-by-term basis.
Immediately, one can
see that (aside from induction, discussed below),
an inclusion of atomic-level anisotropy greatly improves the description of
each energy component. Unless otherwise stated, here we report results
for a\rmse and
dimer-specific fits, though similar values are obtained for overall \rmse and
for transferable fits.
Compared to \isoff, exchange errors in \mastiff
are reduced by 47\%. Electrostatic errors are reduced by an even larger
60\%. By evaluating the ratio of electrostatic errors between different
models, we find that aRMSE(Aniso-Iso)/aRMSE(Iso-Iso) = 0.64
and aRMSE(\mastiff)/aRMSE(Aniso-Iso) = 0.62, suggesting
that \emph{both} higher-order multipoles and anisotropic charge penetration
terms are necessarily to obtain an accurate description of the DFT-SAPT
electrostatic energy. Finally, via an inclusion of dispersion anisotropy,
a\rmse for dispersion are reduced by a significant
65\%. 
%% Especially for dispersion, some of this error reduction may simply be due to
%% the increased number of free parameters associated with anisotropic atom types.
%% (For dispersion, no free parameters are fit for isotropic atom types, while up
%% to three parameters are fit for anisotropic atom types. For 
%% functional forms describing penetration, by contrast, one free parameter is
%% fit for isotropic atom types, while up to four free parameters are fit for
%% anisotropic interactions). If we relax the constraint from
%% \eqref{eq:ff_details} that $\Adisp{} = 1$, thus fitting at least one free
%% parameter per isotropic atom type, we find a somewhat smaller improvement factor for
%% dispersion a\rmse of 47\%, which is more similar to what was observed for the
%% exchange energy. Regardless, it is qualitatively clear that dispersion
%% anisotropy plays an imporant role in determining overall force field accuracy.

Though the trends for exchange, electrostatics, and dispersion universally
suggest the importance of including atomic-level anisotropy, trends for terms
describing the physics of polarization and charge-transfer (represented in
DFT-SAPT by induction and
\dhf) are less encouraging. On the one hand, including higher-order multipoles
substantially lowers \rmse for induction, with aRMSE(Aniso-Iso)/aRMSE(Iso-Iso) = 0.70. 
Because both \isoff and \isaff use isotropic polarizabilities, and because the
induction energy fundamentally depends only on the polarizabilities and the
static electric field, this
result is clearly due to an improved treatment of the static electric field
via anisotropy of the multipolar electrostatics.
Once again, this
suggests that an 
anisotropic treatment of long-range electrostatics is crucial for accurate
force field development. On the other hand, our functional form for anisotropic short-range
induction (\eqref{eq:ff_form} and \eqref{eq:v_aniso}) yields no improvement in the
induction \rmse, with aRMSE(\mastiff)/aRMSE(Aniso-Iso) = 0.97. 
This observed lack of improvement is likely due to a combination of
factors. First, and perhaps most importantly,
we have chosen in this work to use isotropically-averaged dipole
polarizabilities, but as with
electrostatics, anisotropy and higher-order terms have been shown to be important in in the multipole
expansion of atomic dipole polarizabilities. 
\cite{Stone2007,Misquitta2007a,Misquitta2008b,Misquitta2016,Harder2006}
Second, and though probably a smaller source of error,
it is also unclear how to optimally
model the distance dependence of the induction energy at short
intermolecular separations, where penetration and charge-transfer effects
become important and the long-range polarization terms must be damped. 
\cite{VanVleet2016,Liu2017,Misquitta2013,Thole1981} 
Given that the more elaborate short-range form of the 
\mastiff induction model does not result in a tangible improvement, it is quite possible
that alternative formulations are required for an accurate treatment of highly
anisotropic induction.

    %% %%%%%%%%%%%% H2O Comparison %%%%%%%%%%%%%
    %% \begin{figure}[ht]
    %% \includegraphics[width=0.9\textwidth]{figures/h2o_h2o_comparison.pdf}
    %% \caption{
    %%     Force field fits for the water dimer using \isoff (purple), \isaff (orange,
    %%     and \mastiff (green).
    %%     Fits for the total energy are displayed along
    %%     with an inset of the attractive regions. The solid $y=x$ line 
    %%     indicates perfect agreement between reference energies and each force field,
    %%     while dotted lines represent $\pm1$ \kjmol error bounds. \rmse and a\rmse as in
    %%     the main text.
    %%         }
    %% \label{fig:h2o}
    %% \end{figure}
    %% %%%%%%%%%%%% H2O Comparison %%%%%%%%%%%%%

To further analyze the effects of anisotropy on a molecule-by-molecule basis, we have calculated
`improvement ratios', defined as 
aRMSE(Iso-Iso)/aRMSE(\mastiff),
for each energy component and for
each homomonomeric species in the test set, results for which are shown in
\tabref{tab:ratios}. 
(Improvement ratios for heteromonomeric species are given in the \si.) 
The most striking observation from the data presented in \tabref{tab:ratios} is that
the improvement ratios vary considerably with molecule. For example, with water
the a\rmse is improved by an order of magnitude when anisotropy is included. On
the other hand, no improvement is seen for hydrocarbons such as ethane and
methane (also see the \si). Consequently, anisotropy in the short-range expansions
may be necessary for only some atoms types (see \secref{sec:conclusions}). 
In line with chemical intuition, we have found
anisotropy to be particularly important for 
heteroatoms, $\pi$-bonded atoms,
and all hydrogens bonded to anisotropic heavy atoms. 
Appealingly, this distinction between anisotropic and
isotropic atom types 
simplifies force field parameterization and can enable more efficient molecular
simulation (via a more cost-effective treatment of multipolar electrostatics) without sacrificing force field accuracy.
Note that the current empirically-determined
definitions of anisotropic atom types
match both chemical intuition and the more quantitative measures
of atomic anisotropy proposed by other groups.\cite{Kramer2014,Wheatley2012}

    %%%%%%%%%%%% Error Ratios %%%%%%%%%%%%%%%
    \begin{table}[ht]
    \includegraphics[width=0.9\textwidth]{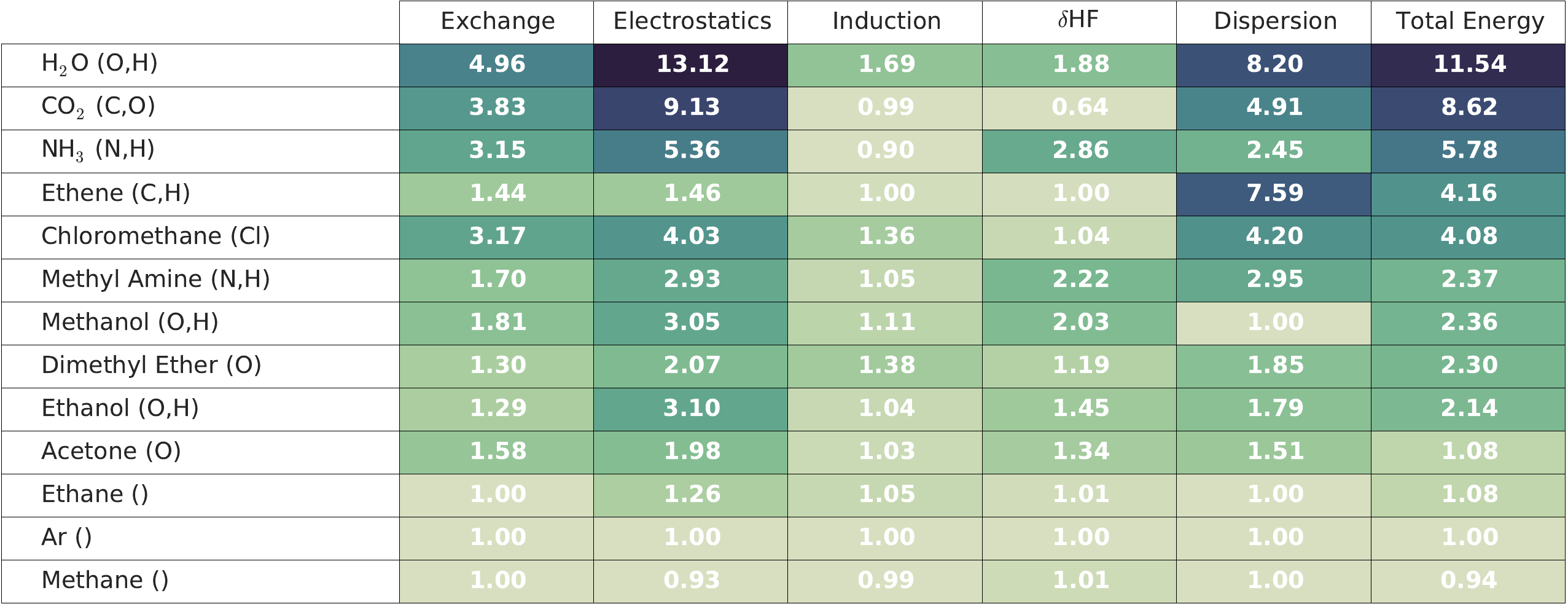}
    \caption{
`Improvement Ratios' for each homomonomeric species in the
91 dimer test set. For each dimer and energy component, the improvement ratio
is calculated as the ratio of a\rmse between \isoff and \mastiff; values
greater than 1 indicate decreased errors in the anisotropic model. Entries have
been ordered according to the improvement ratio for the total energy.
            }
    \label{tab:ratios}
    \end{table}
    %%%%%%%%%%%% Error Ratios %%%%%%%%%%%%%%%

In general, the ordering of improvement ratios for exchange,
electrostatics, dispersion, and the total energies (but not induction, see
above) are reasonably correlated. 
Physically speaking, all atomically-anisotropic
interactions arise from the same source (atomically-anisotropic
electron densities), and so the observed correlation is perhaps to be expected.
Nevertheless, there are some exceptions to this trend.
For ethene, relatively modest improvement ratios (roughly 1.4) are seen for exchange and
electrostatics, whereas dispersion shows a much greater improvement ratio of
7.6. Since ethene homomonomeric interactions are dispersion-dominated, the
improvement ratio for the total energy then roughly corresponds to that of dispersion.
For acetone, there is strong correlation between the improvement ratios for
exchange, electrostatics, and dispersion, which might lead one to suspect that
the total energy improvement ratio would also be around 1.5-2.0. Nevertheless, for this molecule, the
isotropic model benefits from error cancellation between energy components,
and the total energy a\rmse between isotropic and anisotropic models are
rather similar. 

\emph{Crucially, 
our results show that multipolar electrostatics is certainly not the exclusive, nor even
always the dominant, source of atomic anisotropy.}
Indeed, for molecules like ethene, multipolar
anisotropy in the electrostatic model is relatively unimportant, whereas dispersion anisotropy is
essential for accurately modeling the $\pi$ interactions.
Thus, in general, multipolar electrostatic, dispersion, and/or short-range
anisotropies must be all accounted for in order to obtain accurate intermolecular
models.

\end{subsection}
\begin{subsection}{Transferability: Comparison to DFT-SAPT}
\label{sec:transferability}

From the above results it is clear that, when explicitly parameterized,
inclusion of anisotropy can greatly enhance the accuracy of an intermolecular
potential. Nevertheless, for standard force field development, force field
parameters must be \emph{transferable} in order to be
useful in the accurate prediction of intermolecular
interactions in new chemical and/or physical environments. Indeed, in comparing
simpler models to ones that
introduce additional complexity, 
there is an ever-present danger that any accuracy
gains from the more complex functional form are simply due to
over-fitting or error cancellation,\cite{Hawkins2004} ultimately resulting in a
model with poor predictive ability and limited transferability. 

We have previously shown how, with models
similar to \isoff\cite{McDaniel2013,Schmidt2015} or
\isaff,\cite{VanVleet2016} 
it is possible to generate transferable potentials with
applicability to a broad range of chemical and physical
environments.\cite{Schmidt2015} 
This transferability has been
attributed to a
combination of the physically-meaningful energy decomposition of DFT-SAPT,
parameterization on a component-by-component basis (rather than to the
total energy), and the use of physically-motivated functional forms and
parameters.
\cite{VanVleet2016,McDaniel2013,Schmidt2015}
\mastiff largely shares this philosophy of force field development, and so we
might also expect it to be transferable to heteromonomeric dimers. 
Indeed, the long-range multipolar electrostatic model is rigorously
transferable, as are the isotropic long-range induction and dispersion
coefficients used in the force field.\cite{stone2013theory,Stone2007}
However,
the \emph{overall} transferability of \mastiff cannot be taken for granted because of the specific way
in which we have incorporated non-electrostatic anisotropic effects. First, we have relied on several
separability ansatzes (\eqref{eq:separable_anisotropic_repulsion} and
\eqref{eq:aij}), and second, in doing so we have implicitly neglected
potentially important interaction functions that depend on the relative
orientation between monomers (see \appendixref{sec:appendix}). Both of these assumptions may affect the
transferability of the resulting force field.

To assess the transferability of the \mastiff model, we analyze the extent to
which parameters developed for the homomonomeric systems can be used, without
modification, to describe the interactions of the mixed dimers. Such an out-of-sample
prediction, which is easily accomplished with out test set, is a direct
measure of the extent to which our pair potentials can be applied to new
chemical environments. For these transferable fits, parameters were fit to the
13 homomonomeric systems, and the combination rules shown in
\eqref{eq:ff_form} were used
to generate force fields for the remaining heteromonomeric systems. Thus, with these
transferable fits we have essentially generated 78,000 predictions from fits
to 13,000 data points. \rmse and a\rmse for these fits are shown in
\figref{fig:rmse}, and we treat
relative differences between these quantities for the `dimer-specific' and
`transferable' fits as a measure of the extent of transferability for each force
field methodology.

Remarkably, all three force fields --- Iso-Iso, Aniso-Iso, and MASTIFF --- perform
similarly for the dimer-specific and transferable fits, both for the
individual interaction energy components and for the total interaction
energy. The degree of transferability of the \mastiff model is very
encouraging,
and indicates that the manner in which we have chosen to include the
anisotropy is meaningful and does not lead to overfitting, but rather
increases the accuracy of the intermolecular potentials for both in-sample and
out-of-sample systems.

\end{subsection}
\begin{subsection}{Accuracy: Second Virials}
\label{sec:virials}

Having compared our various force fields methodologies against DFT-SAPT, we
now turn our focus to our secondary goal in this work, that of evaluating the
extent to which our anisotropic force field methodology can be used to more
accurately simulate experimental properties.
To this end, we begin by benchmarking our force fields against
experimental second
virial coefficients, 
which offer a direct experimental measure
of the pair potential ($E_2$) without the complication of many-body effects (which
will be discussed in \cref{sec:co2}).
Notably, comparisons to experimental second virial coefficients depend,
not only on the quality of a
force field (as measured in \cref{sec:accuracy}), but also on the accuracy of the benchmark electronic structure
theory used to fit the force field. 
Consequently, and so as to evaluate possible inaccuracies in our 
DFT-SAPT/aVTZ+m\cite{VanVleet2016} benchmark energies,
we have also parameterized
models
with respect to
CCSD(T)-F12a/\avtzm,
a level of theory which serves as a computationally affordable yet accurate prediction of
the CCSD(T)/CBS limit.\cite{Knizia2009,Kalugina2014}
We refer to these coupled cluster-based
models with a -CC suffix, e.g. \mastiff-CC; and details of the refitting
procedure (which minimally effect the dispersion model)
can be found earlier in \secref{sec:methods}. 
Thus, aside from quantum effects (which
are negligible for \co\cite{Bukowski1999a} and well-benchmarked for \ho\cite{Babin2013}), our second virial
predictions should offer a fairly direct comparison between different models,
levels of electronic structure theory,
and experiment.

Using both our original and -CC potentials, we have calculated second virial
coefficients for each \isoffcc, \isaffcc, \mastiff, and
\mastiff-CC, and for the following systems:
\ho (\figref{fig:h2o_virial}), 
\nh (\figref{fig:nh3_virial}), 
\cl (\figref{fig:chloromethane_virial}), and
\co (\figref{fig:co2_virial}).
%
%
    %%%%%%%%%%%% H2O Comparison %%%%%%%%%%%%%
    \begin{figure}[ht]
    \includegraphics[width=0.9\textwidth]{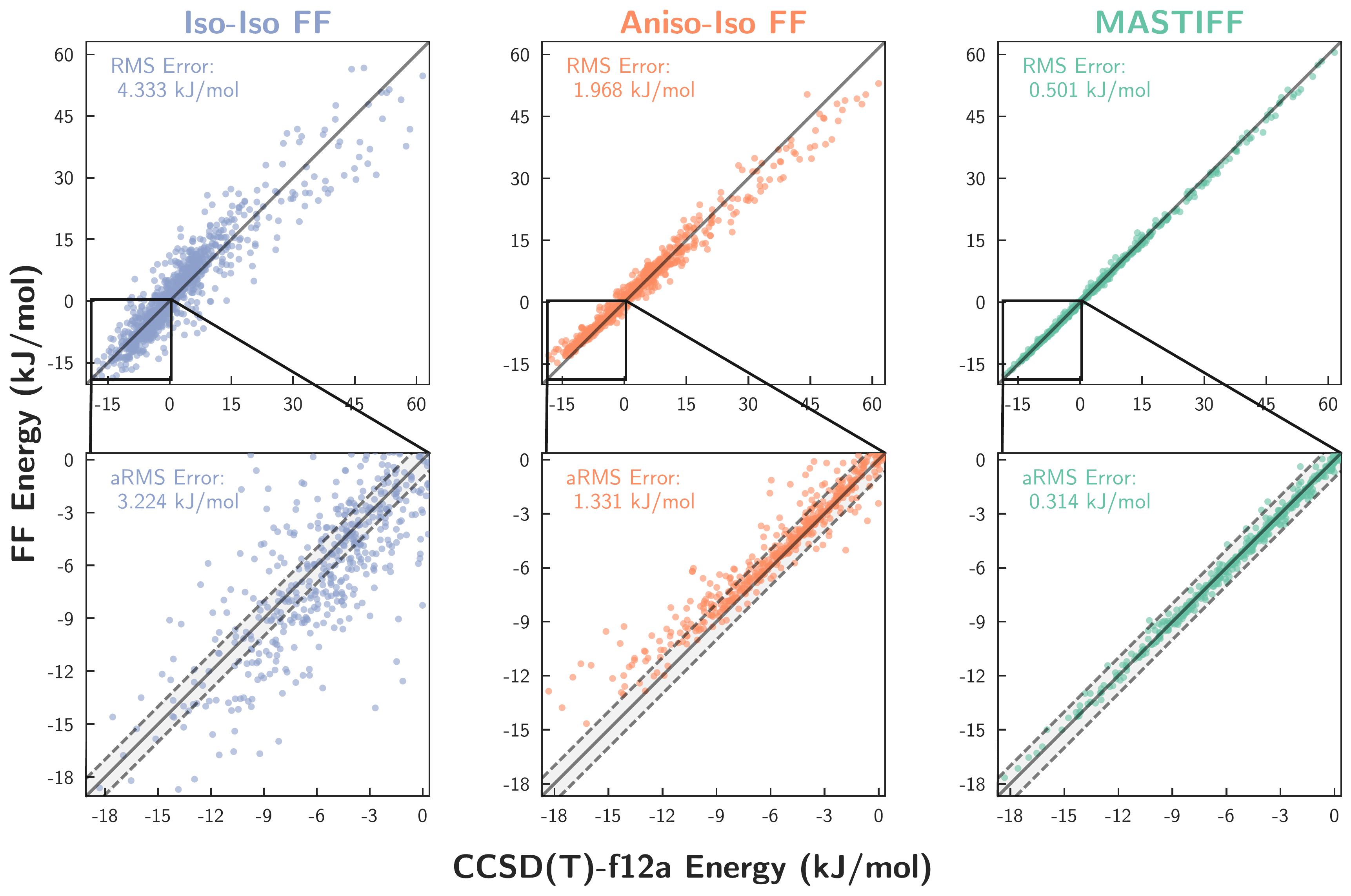}
    \includegraphics[width=0.9\textwidth]{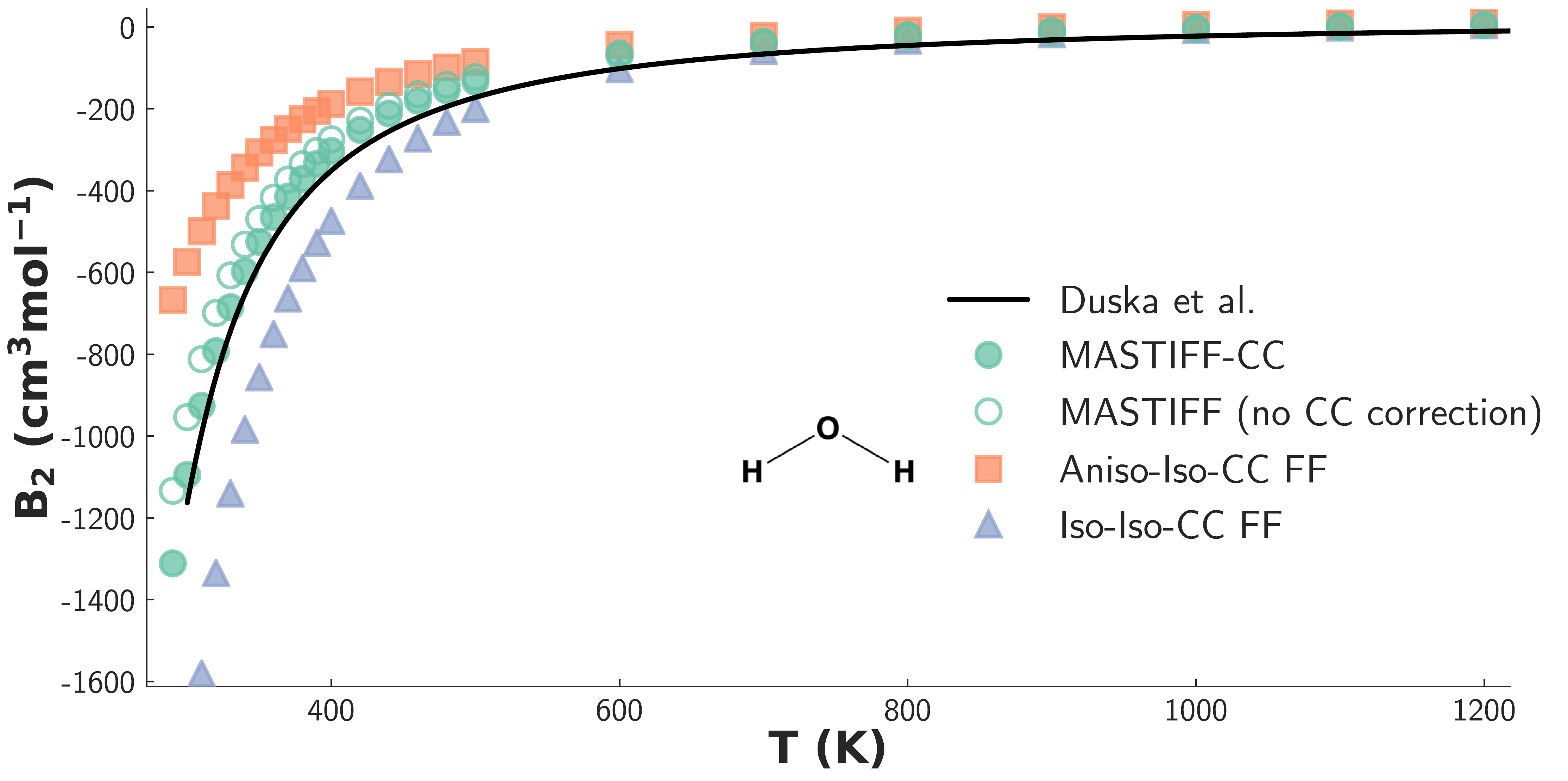}
    \caption{
        (Top) Force field fits for the water dimer using the \isoffcc
(purple), \isaffcc
(orange), and \mastiff-CC (green) methodologies. 
The $y=x$ line indicates perfect agreement between reference
CCSD(T)-F12a
energies and each force field, while shaded gray areas represent points within
$\pm1$ \kjmol agreement of the benchmark. \rmse and a\rmse are as described in
the main text.
(Bottom) Classical second virials for water, with 
        experimental data (black line) taken from
            \citen{Duska2013}. 
        Note that some data points from \isoff extend below the plot area.
            }
    \label{fig:h2o_virial}
    \end{figure}
    %%%%%%%%%%%% H2O Comparison %%%%%%%%%%%%%
    %%%%%%%%%%%% H2O Comparison %%%%%%%%%%%%%
    \begin{figure}[ht]
    \includegraphics[width=0.9\textwidth]{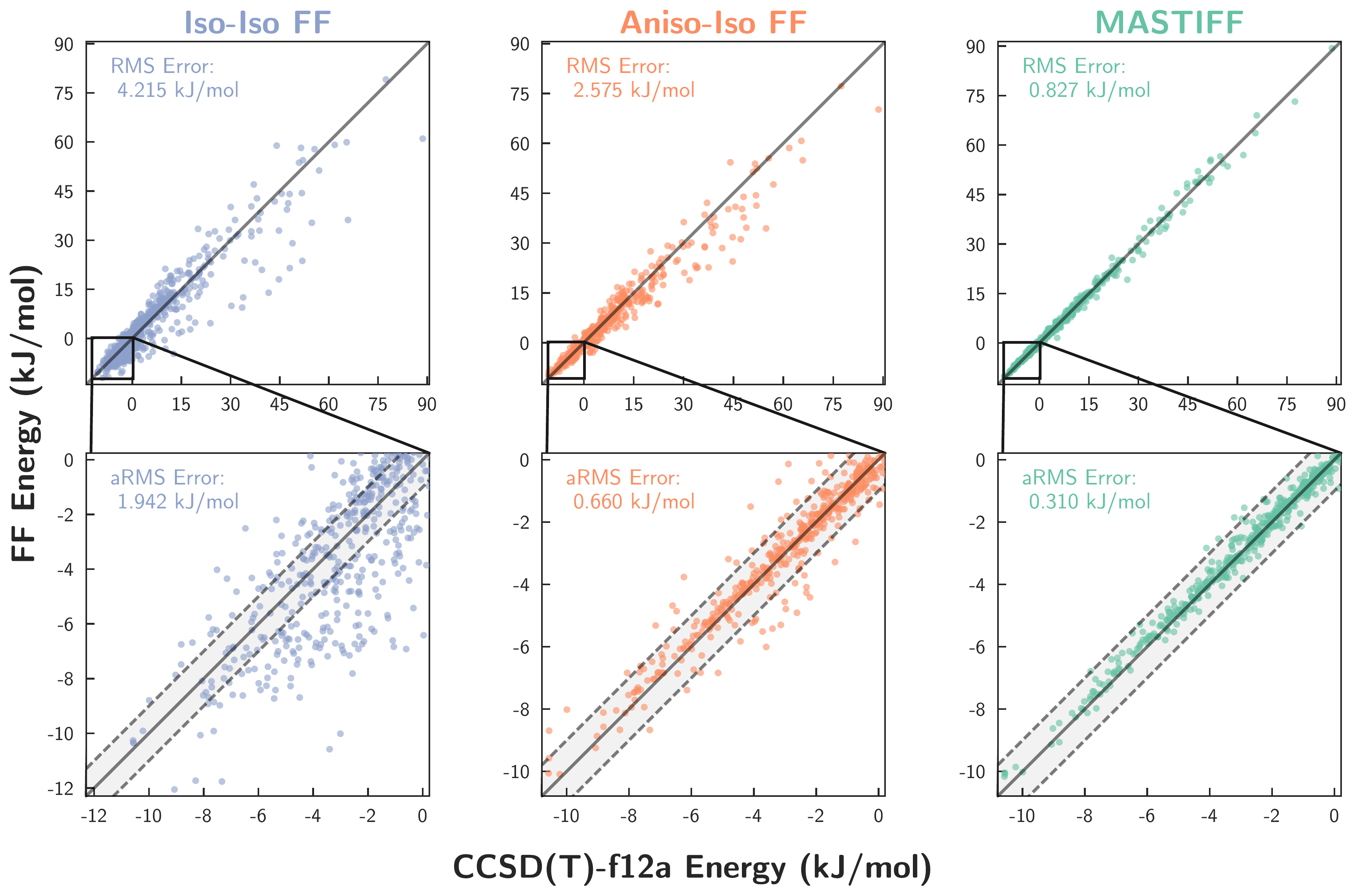}
    \includegraphics[width=0.9\textwidth]{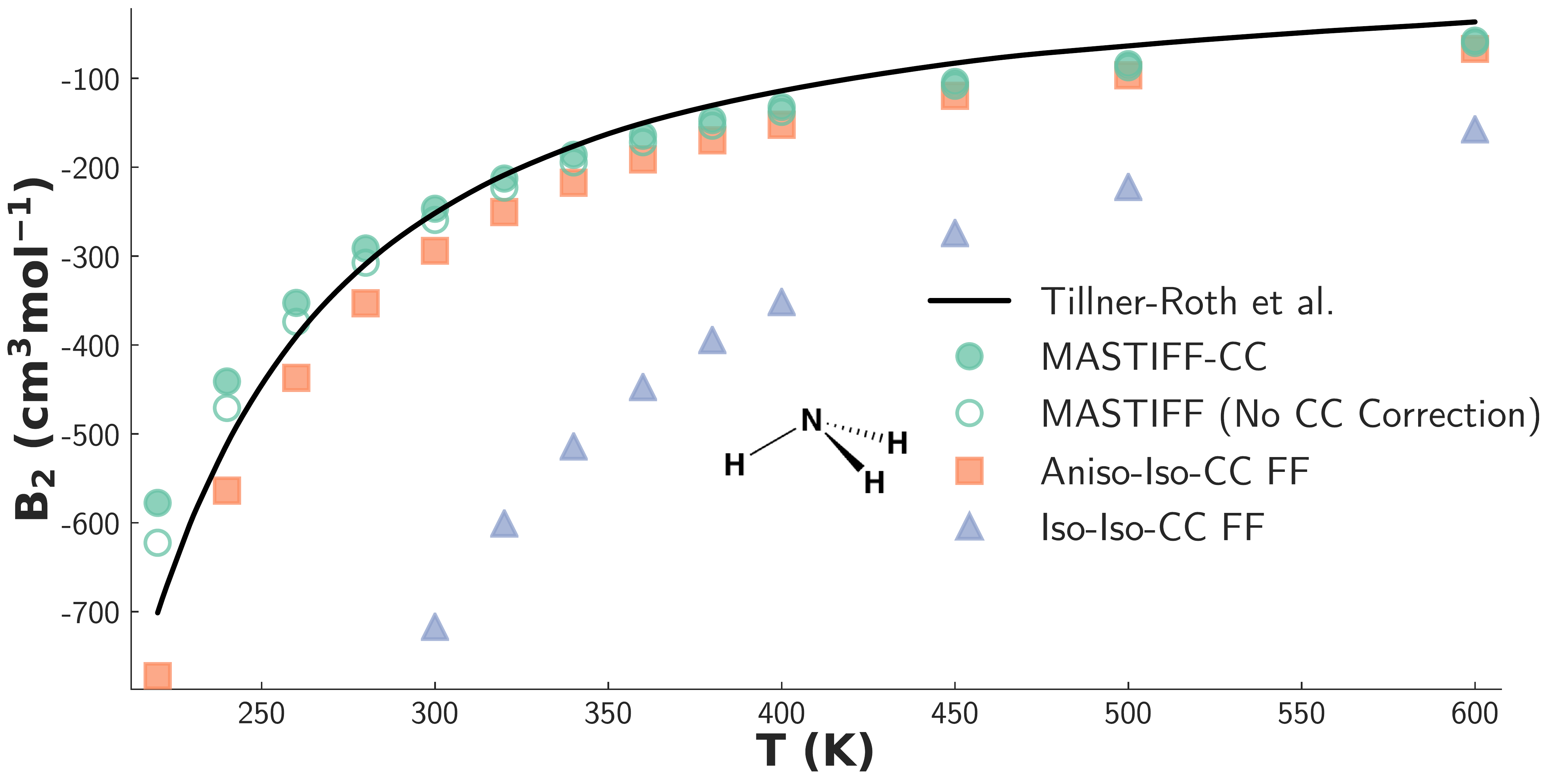}
    \caption{
        Force field fits and classical second virials for ammonia, as in
\cref{fig:h2o_virial}, but with experimental data taken from
            \citen{tillner1993neue}.
            }
    \label{fig:nh3_virial}
    \end{figure}
    %%%%%%%%%%%% H2O Comparison %%%%%%%%%%%%%
    %%%%%%%%%%%% H2O Comparison %%%%%%%%%%%%%
    \begin{figure}[ht]
    \includegraphics[width=0.9\textwidth]{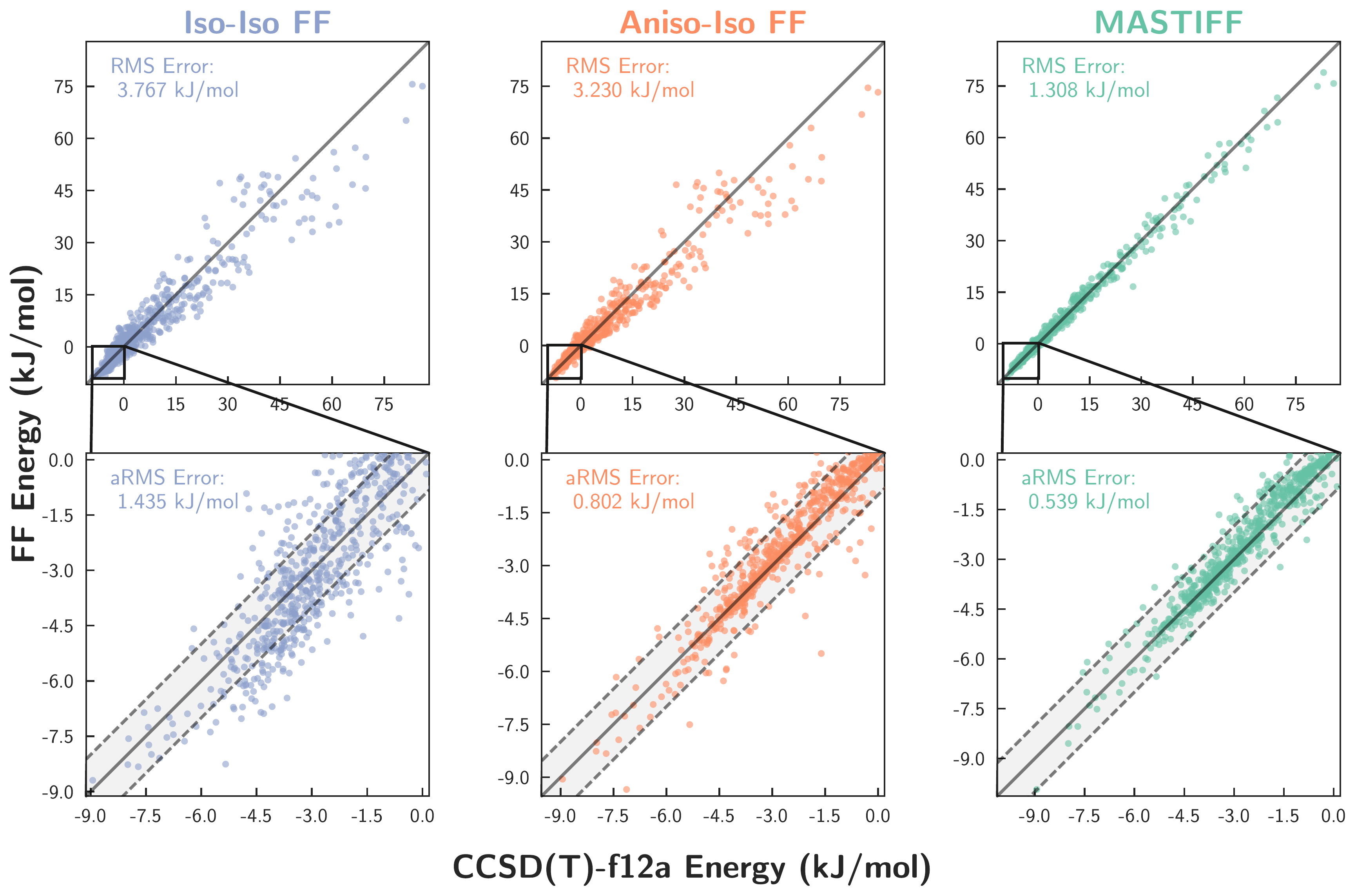}
    \includegraphics[width=0.9\textwidth]{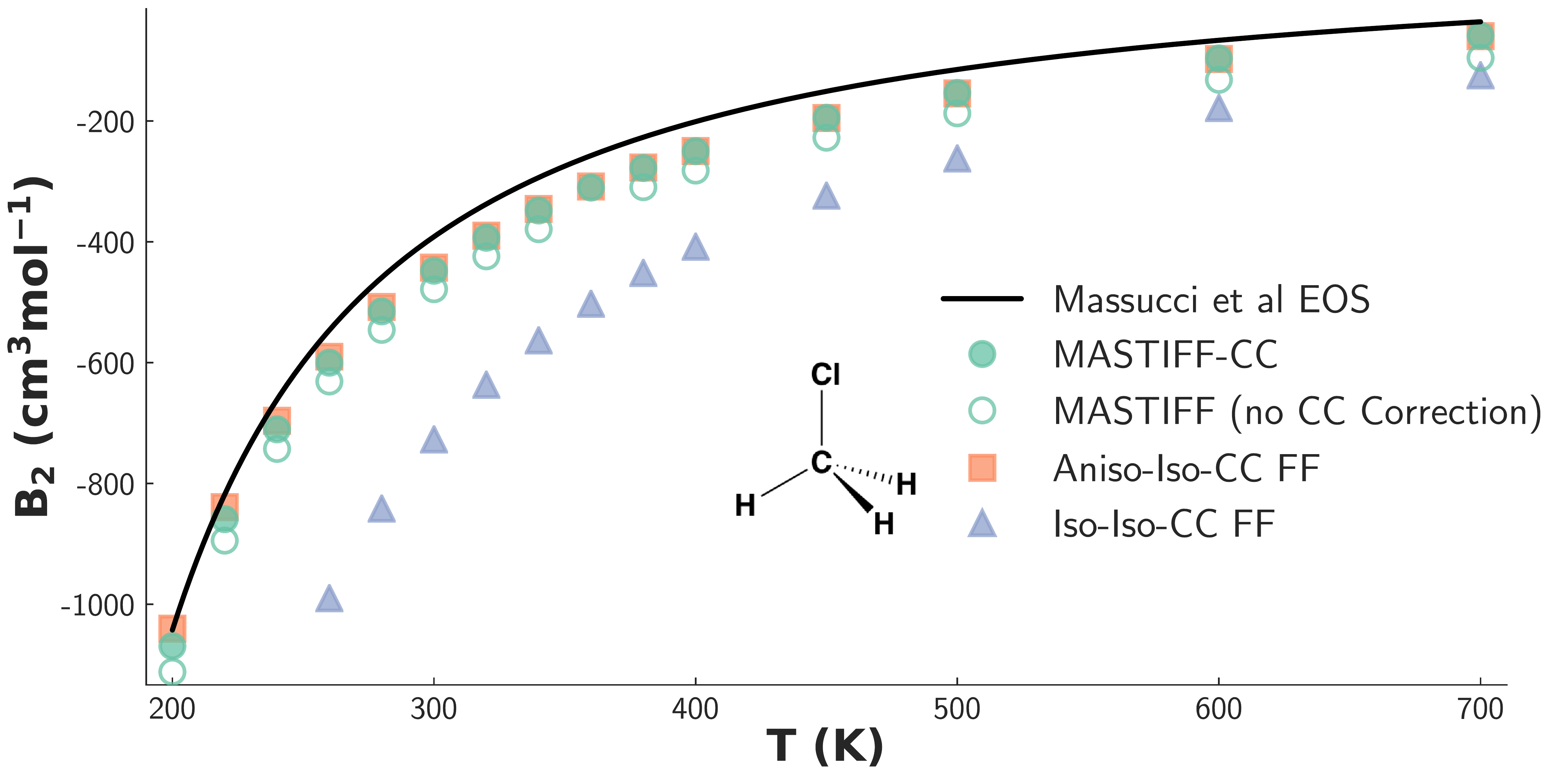}
    \caption{
        Force field fits and classical second virials for chloromethane, as in
\cref{fig:h2o_virial}, but with experimental data from
        the experimental equation of
        state (EOS) given in \citen{Massucci1998}.
            }
    \label{fig:chloromethane_virial}
    \end{figure}
    %%%%%%%%%%%% H2O Comparison %%%%%%%%%%%%%
    %%%%%%%%%%%% Virial for CO2 %%%%%%%%%%%%%
    \begin{figure}[ht]
    \includegraphics[width=0.9\textwidth]{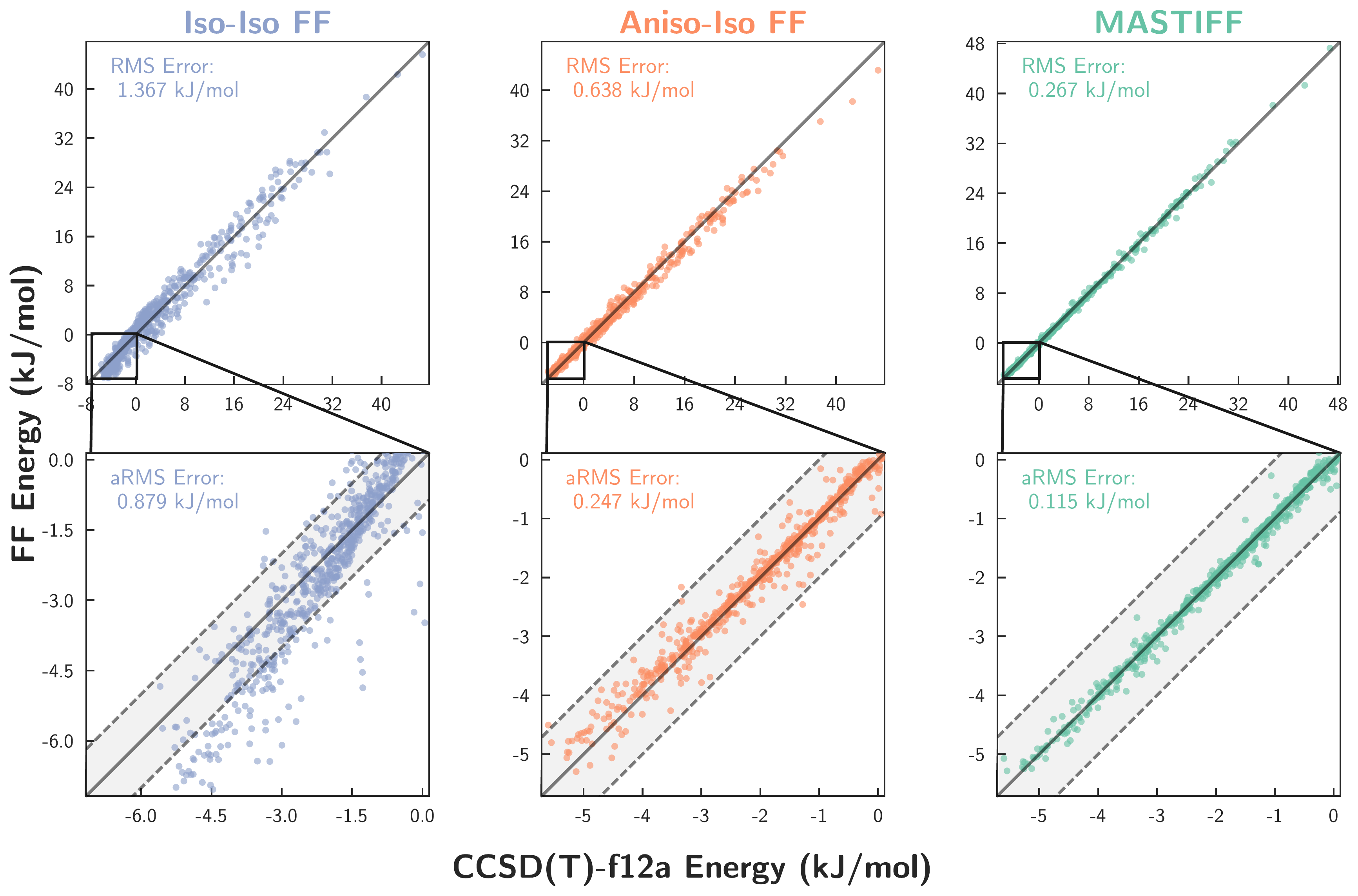}
    \includegraphics[width=0.9\textwidth]{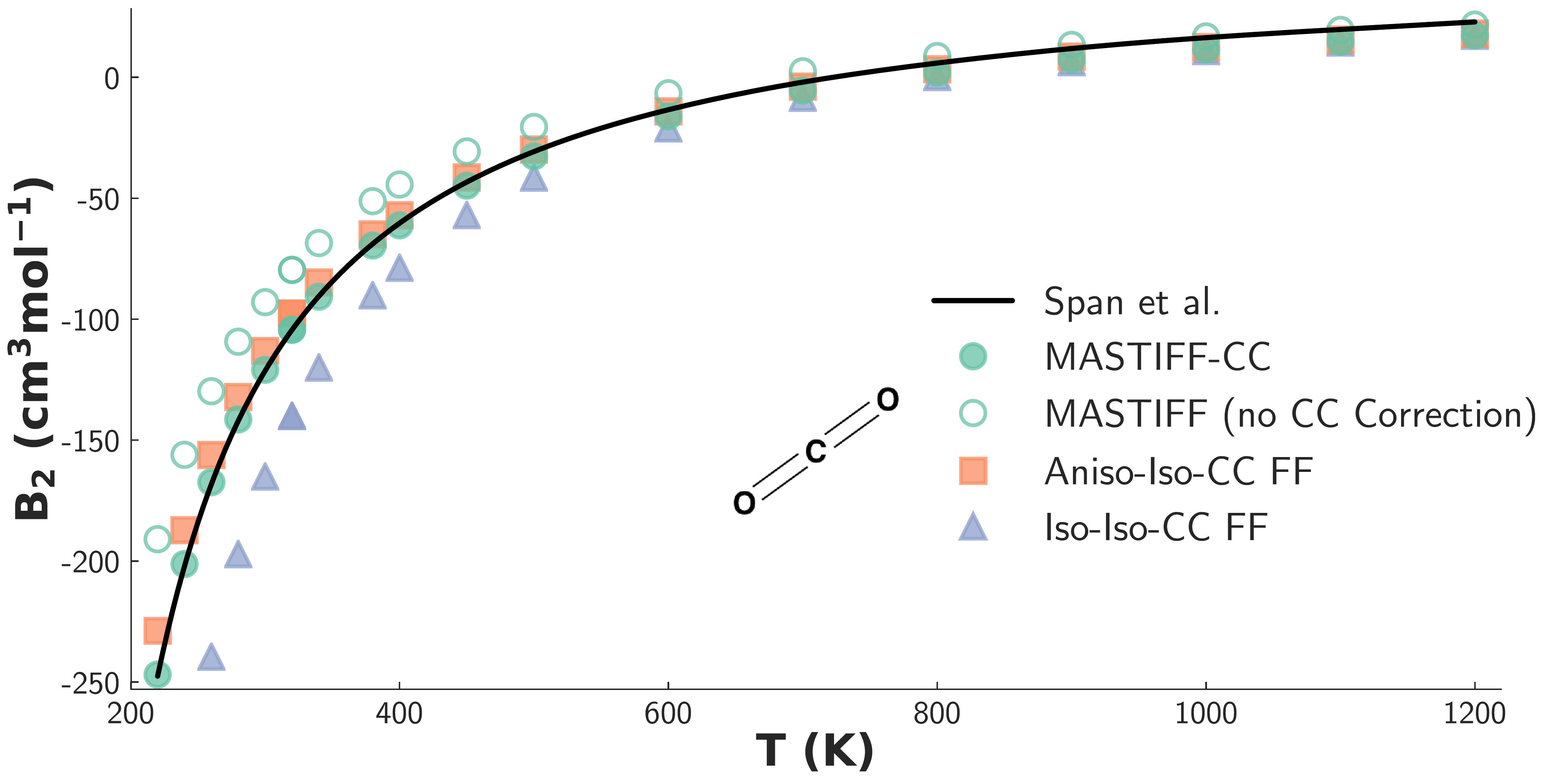}
    \caption{
        Force field fits and classical second virials for \co, as in
\cref{fig:h2o_virial}, but with experimental data taken from
        \citen{Span1996}.
            }
    \label{fig:co2_virial}
    \end{figure}
    %%%%%%%%%%%% Virial for CO2 %%%%%%%%%%%%%
%
Immediately, we observe that the effect of the coupled cluster (-CC)
correction is minimal (compared to differences in force field methodologies)
for most systems, with the exception of \co, where DFT-SAPT exhibits
modest deficiencies with respect to CCSD(T)-f12a (see \si and
\citen{Kalugina2014}). 
Furthermore, we find
that 
that the \mastiff (and especially \mastiff-CC)
methodologies predict virial coefficients
which closely corresponds to experimental data. 
In general, the \isoffcc 
predictions are much worse than their \mastiff-CC or \isaffcc counterparts, 
suggesting that
an accurate treatment of long-range electrostatics is essential to obtain
accurate virial coefficients.
Finally, though \isaff-CC gives equally good predictions for some systems
(notably \cl) compared to the \mastiff-CC method, 
virial coefficients for other systems (especially \ho) are less
accurate, suggesting that dispersion and short-range
anisotropies are also important in many systems for the accurate prediction of
virial coefficients. 

In general, and given the range of systems tested (\co dimer interactions are
dispersion dominated, while \cl, \nh, and \ho have relatively larger
electrostatic and polarization contributions), these second virial calculations
suggest that, when fit to gold-standard electronic structure theories, our
anisotropic force field methodology offers an improved strategy for
developing quantitatively accurate pair potentials.

\clearpage
\end{subsection}
\begin{subsection}{Accuracy: Condensed Phase Properties of CO$_2$}
\label{sec:co2}

A major goal for standard force fields is that they be capable of accurately simulating
bulk properties. To this end, we require not only an
accurate pair potential, but also (in many cases) a proper treatment of polarization
and other many-body effects. 
So as to provide a first example of how
the \mastiff methodology might be used as the pair potential in a complete,
many-body force field useful
for condensed
phase simulation, 
here we have developed and tested a
force field for \co which includes both pairwise additive and many-body
effects.
Based on its accuracy in predicting second virial coefficients, we use the \mastiff-CC
potential from \cref{sec:virials} to describe both the pairwise potential and the many-body induction.
Yet, non-inductive many-body effects have been shown to be
important for \co,
\cite{Yu2012b,Hellmann2017,Oakley2009a,Desgranges2015} 
and so we have additionally developed and tested a model for three-body dispersion 
based on the three-body dispersion potential developed by
\citeauthor{Oakley2009a} (see \secref{sec:methods}).  Three-body exchange
effects are not accounted for in our model, however 
prior work shows they are very small under the conditions studied
here.\cite{Yu2012b}
Using the various \co models described above, 
% (i.e. with and without inclusion of three-body dispersion effects),
we
have run bulk simulations for a rigid model of \co over a variety
of vapor, liquid, supercritical, and solid phase points.
Density predictions for the vapor, liquid, and supercritical phases of \co are
shown in \tabref{tab:densities} and in the \si (Tables S3 and S4), and enthalpies of
sublimation and vaporization are shown in \tabref{tab:deltah} and Table S5. 
Simulations with a flexible \co model yielded similar results, which are also
given in the \si. 

As anticipated from prior work,\cite{Yu2012a} complete neglect of three-body
dispersion (Table S3) leads
to an overestimation of the density at all phase points studied, particularly
in the denser liquid phases. Though not surprising, this result underscores
the importance of including many-body effects (at least for \co) when
developing highly accurate ab initio force
fields.\cite{Yu2012b,McDaniel2014}
Upon including three-body dispersion effects,
however,
\mastiff-CC succeeds in reproducing \emph{all} studied experimental properties to within a
few percent (see \cref{tab:densities,tab:deltah}). (As shown in Table S4, \isaffcc
reproduces some, but not all, experimental properties to within this level of
accuracy, and \isoffcc generally has poor quantitative agreement with
experiment.)
Importantly, \mastiff-CC can correctly predict
the
\co sublimation enthalpy, a quantity which critically depends on the lattice energy of the
solid phase. Unlike with liquid or supercritical \co, where
many dimer configurations are sampled,
the
solid consists of only four symmetry-unique
configurations. Consequently, whereas an isotropic
potential 
% (which is in error for particular dimer
% configurations, but can take advantage of error cancellation to be accurate in an average sense) 
might yield good property predictions for the liquid phase via averaging
and/or error cancellation,
it would not be expected to correctly predict the solid phase, where beneficial
error cancellation is unlikely. 
Indeed, most theories (including \isaffcc, \isoffcc, our previously developed SYM-3B
model,\cite{Yu2012b} nearly all popular empirically-developed \co
models,\cite{Perez-Sanchez2013} AMOEBA,\cite{Heit2016a} and many electronic
structure theories\cite{Heit2016a}) struggle to correctly predict the solid phase
properties of \co. For this reason, 
the enthalpy of sublimation is
considered
an extremely stringent test of force field
quality,\cite{Perez-Sanchez2013} 
and the fact that \mastiff-CC can accurately reproduce %our accurate reproduction
this quantity is evidence for both the excellent quality of the many-body \mastiff-CC
potential in specific and of the importance of atomic-level anisotropy in
general.
%% Though more testing is needed to confirm
%% the accuracy of our anisotropic force field for
%% other phase points, our results suggest that, crucially, the
%% \mastiff-CC potential is transferable across the entire phase diagram of
%% molecular \co, and is capable of describing the gas, liquid, supercritical,
%% and solid phases. 
%
Overall, our \co results are a preliminary indication
that,
provided we correctly account for many-body effects, and benchmark against a
gold-standard electronic structure theory, our newly developed anisotropic
methodology may successfully be used as the basis for accurate, `next-generation' force
fields amenable to the molecular simulation of bulk properties in a variety of
phases.

Despite the success of our \mastiff-CC model for \co, it is also worthwhile to
address and understand its minor shortcomings. In particular, we have studied 
representative two- and three-body energies taken from a snapshot of
the liquid at 273.15 K and 100 bar (see Figure S5 in the \si). 
%% For the two-body energies, we have compared
%% against the extremely accurate \citeboth{Kalugina2014} potential, while for
%% three-body energies we have benchmarked against the \citeboth{Hellmann2017}
%% PES. From these results (shown in the \si), 
%% it is clear that our pairwise
%% \mastiff-CC potential is highly accurate for all configurations present in the
%% liquid, with very small \rmse and no systematic error in the
%% potential, such that the total two-body energy is accurate to within 0.05\%
%% compared to the \citeauthor{Kalugina2014} PES. 
%
%% Once again, this result argues
%% strongly for the accuracy and transferability of the \mastiff methodology, and
%% suggests that an inclusion of anisotropy is essential, not only for gas-phase
%% clusters, but also for simulations of the bulk. 
When benchmarked against the accurate PES developed by
\citeauthor{Hellmann2017},\cite{Hellmann2017} the crude three-body
potential utilized above is found to be systematically in error. 
Though some of this error may be due to inaccuracies in the
benchmark potential itself, as compared to coupled-cluster,\cite{Hellmann2017}
most of this error is likely due to
inaccuracies in our model for many-body \co interactions. The
atomically-isotropic treatment of three-body dispersion, neglect of
higher-order dispersion terms,
and neglect of explicit three-body exchange
may all contribute to this error, and
an improved model for many-body \co interactions will be the subject of future
research. Indeed, it is well-known that the density can be extremely sensitive
to the treatment of many-body effects,\cite{Desgranges2015} and it is highly
probable that an improved many-body model would reduce the already small
errors observed in our \mastiff-CC predictions.
Regardless, (and despite some small residual errors
arising from the simplified treatment of many-body effects) it appears that
the \mastiff-CC
methodology yields an extremely accurate two-body force field for \co, 
with broad
applicability across 
a range of experimentally-important phases.
%

%%%%%%%%%%%%%%%%%%%%% Macroscopic Properties %%%%%%%%%%%%%%%%%%%%%%%%%%%%%%%%%%
\begin{table}
%\small
\centering
\renewcommand\arraystretch{1.1}
\begin{tabular}{@{}lccccc@{}}
\hline
\toprule

Phase & T (K) & P (bar) &  Density (g/ml) &  Exp. & \% Error  \\

\midrule
Gas       & 300    &  50  & 0.131   & 0.128 &  2.34  \\
Supercritical & 320    & 140  & 0.728   & 0.703 &  3.56  \\
Liquid    & 300    & 100  & 0.825   & 0.802 &  2.87  \\
Liquid    & 273.15 & 100  & 1.000   & 0.974 &  2.67  \\
\bottomrule
\hline
\end{tabular}
\caption{
    Select densities for \co across a range of experimental conditions.
    Experimental data taken from the EOS of \citen{Span1996}.
    Entries ordered by increasing experimental density.
	}
\label{tab:densities}
\end{table}
%\normalsize
%%%%%%%%%%%%%%%%%%%%% Macroscopic Properties %%%%%%%%%%%%%%%%%%%%%%%%%%%%%%%%%%

%%%%%%%%%%%%%%%%%%%%% Macroscopic Properties %%%%%%%%%%%%%%%%%%%%%%%%%%%%%%%%%%
\begin{table}
%\small
\centering
\renewcommand\arraystretch{1.1}
\begin{tabular}{@{}lccccc@{}}
\hline
\toprule

Phases & T (K) & \deltah (\kjmol) &  Exp. & \% Error  \\

\midrule
s $\to$ g   & 194.76 & $25.0 \pm 0.15$ &  25.2  &  -0.8  \\
l $\to$ g   & 288    & 7.92   &    7.80    &   -1.4 \\
\bottomrule
\hline
\end{tabular}
\caption{
    Enthalpies of vaporization/sublimation for \co at several temperatures. 
    Experimental data taken from the EOS of \citen{Span1996}.
    The uncertainty in the enthalpy of sublimation is due to ambiguity in the theoretical zero-point energy for \co (see
    \secref{sec:methods}. 
	}
\label{tab:deltah}
\end{table}
%\normalsize
%%%%%%%%%%%%%%%%%%%%% Macroscopic Properties %%%%%%%%%%%%%%%%%%%%%%%%%%%%%%%%%%

\end{subsection}

%% Explain MSE Fits
%%     Exchange:
%%     Electrostatics:
%%     Dispersion: 
%%     Total Energy:

\end{section}
%%%%%%%%%%%%%%%%%%%%%%%%%%%%%%%%%% Results %%%%%%%%%%%%%%%%%%%%%%%%%%%%%%%%%%%%%%%%

%%%%%%%%%%%%%%%%%%%%%%%%%%%%% Conclusions %%%%%%%%%%%%%%%%%%%%%%%%%%%%%%%%%%%%%%%%%
\begin{section}{Conclusions and Future Work}
\label{sec:conclusions}

We have developed a comprehensive methodology for modeling 
atomic-level anisotropy in standard intermolecular force fields. 
Via a simple extension to
standard isotropic force
fields,\cite{VanVleet2016}
%% and by accounting for this anisotropy in our models for each
%% electrostatics, exchange-repulsion, induction, and dispersion,
we have 
demonstrated how a computationally-efficient treatment of
atomic-level anisotropy can lead to
significant improvements in models for intermolecular interactions.
Critically, 
and in contrast to popular assumption, an accurate treatment of multipolar electrostatics 
does not \emph{by itself} account for all energetically-important effects of
atomic-level anisotropy.
Rather, our results indicate that the combined anisotropy
of 
dispersion, exchange, and charge penetration 
is of comparable importance to long-range multipolar electrostatics, and must be
comprehensively accounted for in order to obtain
intermolecular
force fields of the highest quality.
In agreement with the more quantitative metrics proposed by others,\cite{Wheatley2012,Kramer2014} 
we have found
a comprehensive model of atomic-level anisotropy to be particularly important for obtaining sub-\kjmol
accuracy when describing molecules with heteroatoms
(particularly ones with exposed lone pairs), carbons in multiple
bonding environments, and hydrogens bound to anisotropic heavy atoms. 
As such, our `\mastiff' methodology show great promise with
respect to both high-quality electronic structure benchmark energies and
experimental property predictions,
all
while maintaining high transferability and ease of implementation in
existing software packages for use in condensed phase
simulation.\cite{Eastman2013}

Nonetheless, several aspects of our current force
field methodology require further improvement and/or study
before our anisotropic \mastiff approach can be used to develop
standard force fields for arbitrary organic and/or
biological systems. 
As an example, 
future work will be required to investigate
how well the \mastiff methodology can be applied to studies of large and/or
non-rigid systems,
though similar isotropic models have previously been shown
to transferably combine 
with intramolecular potentials in order to describe molecular
flexibility.\cite{McDaniel2013} 
Additionally, an improved description of induction effects
will become essential for accurate bulk simulations of highly polarizable
molecules, such as water.  We are currently working to develop improved
models that can describe both long-range anisotropic polarization and
short-range polarization damping, as these aspects of the force field
critically affect both the two- and many-body induction energies and
can account for a sizable fraction of the total interaction energy in
condensed phases.
We anticipate that improved models for molecular flexibility and induction will, in
combination with an accurate description of non-inductive many-body effects,
yield a general approach to force field development that accurately models
arbitrary $N$-body 
intermolecular interactions, in turn
enabling highly accurate, `next-generation' force field development capable of
simulating a wide array of phases and chemical environments.

\end{section}
%%%%%%%%%%%%%%%%%%%%%%%%%%%%% Conclusions %%%%%%%%%%%%%%%%%%%%%%%%%%%%%%%%%%%%%%%%%

%%%%%%%%%%%%%%%%%%%%%%%%%%%%% Acknowledgements %%%%%%%%%%%%%%%%%%%%%%%%%%%%%%%%%%%%
\begin{acknowledgement}

This material is based upon work supported by the National Science
Foundation Graduate Research Fellowship under Grant No. DGE-1256259 and 
by Chemical Sciences, Geosciences and Biosciences
Division, Office of Basic Energy Sciences, Office of Science, U.S. Department
of Energy, under award DE-SC0014059.  
J.R.S is a Camille Dreyfus
Teacher-Scholar. M.V.V. thanks Dr. Ken Jordan, Dr. Greg Beran, Dr. Anthony Stone, and
especially Dr. Jesse McDaniel for
many helpful discussions, and acknowledges Dr. Sarah L. Price and Queen Mary
University of London for travel funding as this work was completed.
Computational resources were provided in part by National Science Foundation
Grant CHE-0840494 and using the computational resources and assistance of the
UW-Madison Center for High Throughput Computing (CHTC) in the Department of
Computer Sciences. The CHTC is supported by UW-Madison, the Advanced Computing
Initiative, the Wisconsin Alumni Research Foundation, the Wisconsin Institutes
for Discovery, and the National Science Foundation, and is an active member of
the Open Science Grid, which is supported by the National Science Foundation
and the U.S. Department of Energy's Office of Science.
Compuational resources were also provided in part by the UW Madison Chemistry
Department cluster Phoenix under grant number CHE-0840494, and by 
the Extreme Science and Engineering Discovery Environment
(XSEDE), which is supported by National Science Foundation grant numbers
TG-CHE120088 and TG-CHE170079. 

\end{acknowledgement}
%%%%%%%%%%%%%%%%%%%%%%%%%%%%% Acknowledgements %%%%%%%%%%%%%%%%%%%%%%%%%%%%%%%%%%%%

%%%%%%%%%%%%%%%%%%%%%%%%%%%%% Supporting Info %%%%%%%%%%%%%%%%%%%%%%%%%%%%%%%%%%%%%
\begin{suppinfo}
\mse values for the 91 dimer test set. 
Improvement ratios for all 91 dimers. 
Local axis definitions for all 13 monomers.
\mastiff parameters for homomonomeric systems.
\mastiff-CC parameters and OpenMM input files for \ho, \co, \nh, and \cl.
\isoff, \isaff, and \mastiff fitting quality for homomonomeric systems.
Representative 2- and 3-body energies taken from liquid \co.
\end{suppinfo}
%%%%%%%%%%%%%%%%%%%%%%%%%%%%% Supporting Info %%%%%%%%%%%%%%%%%%%%%%%%%%%%%%%%%%%%%

%%%%%%%%%%%%%%%%%%%%%%%%%%%%% Appendix %%%%%%%%%%%%%%%%%%%%%%%%%%%%%%%%%%%%%%%%%%%%
\begin{appendices}
%\appendix
\begin{section}{\sfunc and the Motivation for \gij}
\label{sec:appendix}

As shown elsewhere,\cite{Stone1978,Stone1984}
an exact (under the ansatz of
radial and angular separability) model for \gij is given by Stone's $\bar{S}$-functions, 
which form a complete basis set for describing any scalar function which
depends on the relative orientation
between molecules. These \sfunc are given (following Stone's
notation\cite{stone2013theory}) by the formula
% General s-function
\begin{align}
\bar{S}^{k_1 k_2}_{l_1 l_2 j} = 
i^{l_1 - l_2 -j}
\begin{pmatrix} l_1 & l_2 & j \\ 0 & 0 & 0 \end{pmatrix}^{-1}
\sum \limits_{m_1 m_2 m} 
[D^{l_1}_{m_1 k_1}(\Omega_1)]^*
[D^{l_2}_{m_2 k_2}(\Omega_2)]^*
C_{lm}(\theta,\phi)
\begin{pmatrix} l_1 & l_2 & j \\ m_1 & m_2 & m \end{pmatrix}.
\end{align}
The general form of these $\bar{S}$-functions can be quite complicated, and
involve both the Wigner $D$ rotation matrices and Wigner $3j$-symbols (quantities
in parentheses) as well as the degree ($l_1$, $l_2$, and $j$) and order
($m_1$, $m_2$, and $m$ for the global coordinate system, $k_1$ and $k_2$ for the
various local coordinate systems) of the spherical harmonic tensors. Here
subscripts reference either molecule 1 or molecule 2, and subscriptless
quantities refer to the dimer as a whole. 

In order to obtain a functional form for the exchange-repulsion that is
amenable to simple combination rules (a necessary prerequisite for
transferable potentials), we must somehow be able to separate \gij into monomer
contributions. Unfortunately, many of the \sfunc depend on the relative
orientation of the dimer itself, and thus must be excluded in the development
of \emph{transferable} potentials.
Thus as a second ansatz (empirically validated by us in \secref{sec:results} and by
others\cite{Millot1992})
we neglect all contributions from \sfunc that depend
on both local coordinate systems. This leaves us with two sets of \sfunc, namely
\begin{align}
\bar{S}^{k0}_{l0l} = C_{lk}(\theta_i,\phi_i)
\end{align}
and
\begin{align}
\bar{S}^{0k}_{0ll} = C_{lk}(\theta_j,\phi_j)
\end{align}
which are simply the renormalized spherical harmonics (\eqref{eq:sph_harm})
expressed in each of the two local coordinate systems.

Given our truncated expressions for the \sfunc, we now need only extend our
functional form for \fij to incorporate these anisotropic contributions.
We choose, in a manner analogous to literature precedent,
\cite{Stone2007,Mitchell2001,Price2000,Stone1988,Day2003,Torheyden2006,Totton2010,Misquitta2016,Price2010a}
to expand the \Aex{i} and \Aex{j} parameters of \eqref{eq:aij} in terms of a truncated
expansion of \sfunc. 
(In principle, we could also account for anisotropy in the \B parameters of
our model for \fij. However, previous literature suggests that in practice this `hardness'
parameter can often be treated as constant, and we also neglect its possible
anisotropy in this work.) 
Consequently, all short-range anisotropies are modeled in this work
by the expressions given in \eqref{eq:gij} and \eqref{eq:vex}.

In addition to describing exchange-repulsion, \sfunc can also be used to
accurately describe the orientation dependence of long-range electrostatic, 
induction, and dispersion energies.
(See \citens{stone2013theory,Stone2007} for complete details.)
The electrostatic
interaction tensor from \cref{eq:electrostatic_t} can be expressed, in terms of
\sfunc, as\cite{stone2013theory}
\begin{align}
T^{ij}_{tu} \equiv T^{ij}_{l_1,k_1, l_2,k_2} = \begin{pmatrix} l_1
+ l_2 \\ l_1 \end{pmatrix} \bar{S}^{k_1 \ k_2}_{l_1 \ l_2 \ l_1+l_2} \ r_{ij}^{-l_1 - l_2 - 1}
\end{align}
where both the $tu$ or $l_1,k_1,l_2,k_2$ notations label the angular
momentum of the multipole components.
The long-range induction energy is also explicitly dependent on the electrostatic
interaction tensor (and hence implicitly dependent on the
\sfunc),\cite{stone2013theory}
\begin{align}
\vind = \frac12 \sum\limits_{I} \sum\limits_{I\ne J} \Delta Q^i_t T^{ij}_{tu}
Q^{j}_u, 
\end{align}
with $\Delta Q$ and $Q$ defining the induced and permanent multipoles,
respectively, and $I$ and $J$ representing individual molecules.
Lastly, the orientation dependence of the long-range dispersion is
accruately described by
the formula\cite{Stone2007}
\begin{align}
\vdisp{} = - \frac{1}{2\pi} \sum\limits_{iji'j'} \sum\limits_{tut'u'}
T^{ij}_{tu} T^{i'j'}_{t'u'} \int\limits^{\infty}_{0}
\alpha^{ii'}_{tt'}(\text{i}\nu)
\alpha^{jj'}_{uu'} (\text{i}\nu) d\nu,
\end{align}
where the primes describe the
response of the local polarizability ($\alpha^{ii'}$)
at site $i'$ to a perturbation at $i$, and the integration is carried out over all
imaginary frequencies $\text{i}\nu$.

%% Finally, and as with exchange-repulsion, molecular anisotropic long-range
%% dispersion coefficients can be determined from an \sfunc expansion of the form
%% %
%% \begin{align}
%% \vdisp = - \sum\limits_{l_A l'_A l_B l'_B}
%%            \sum\limits_{L_A L_B J K_A K_B}
%%             \bar{C}_n(L_A L_B J; K_A K_B)R^{-n} S^{K_A K_B}_{L_A L_B J}
%% \end{align} 
%% %

\end{section}
\end{appendices}
%%%%%%%%%%%%%%%%%%%%%%%%%%%%% Appendix %%%%%%%%%%%%%%%%%%%%%%%%%%%%%%%%%%%%%%%%%%%%

\clearpage
\singlespacing

\renewcommand{\baselinestretch}{1}

\bibliography{library}
%\bibliography{library,misquitta}

%%%%%%%%%% TOC Graphic %%%%%%%%%%%%%%%
%% \clearpage
%% \begin{section}{TOC Graphic}
%% \includegraphics[width=0.9\textwidth]{toc/toc.pdf}
%% \end{section}

\end{document}